\documentclass[journal]{IEEEtran}

\usepackage{amsfonts}
\usepackage{amssymb}
\usepackage{stfloats}
\usepackage{cite}
\usepackage{graphicx}
\usepackage{epstopdf}
\usepackage{psfrag}
\usepackage{subfigure}
\usepackage{amsmath}
\usepackage{array}
\usepackage{multirow}
\usepackage{color}
\usepackage{booktabs}
\usepackage{url}
\usepackage{enumitem}
\usepackage{lipsum}

\usepackage{marvosym} 
\usepackage{hyperref}
\usepackage{makecell}
\usepackage{bm}

\usepackage[linesnumbered,ruled,vlined]{algorithm2e}

\newtheorem{Thm}{Theorem}
\newtheorem{Lem}{Lemma}

\newtheorem{Def}{Definition}

\newtheorem{Prob}{Problem}

\newtheorem{Prop}{Proposition}
\newtheorem{Assump}{Assumption}

\begin{document}

\title{Green Wireless Network Scaling for Joint Deployment: Multi-BSs or Multi-RISs?}

\author{Tao~Yu, Simin~Wang,  Shunqing~Zhang,~\IEEEmembership{Senior Member, IEEE}, Mingyao~Cui, Kaibin~Huang,~\IEEEmembership{Fellow, IEEE}, Wen~Chen,~\IEEEmembership{Senior Member, IEEE}, QingQing~Wu,~\IEEEmembership{Senior Member, IEEE}, Jihong~Li and Kaixuan~Huang 
\thanks{This work was supported by the National Key Research and Development Program of China under Grants 2022YFB2902304, the Science and Technology Commission Foundation of Shanghai under Grant 24DP1500500, the National Natural Science Foundation of China under Grant 62571307 and the Science and Technology Commission Foundation of Shanghai under Grants 24DP1500703.}
\thanks{Tao Yu, Mingyao Cui and Kaibin Huang are with the Department of Electrical and
Electronic Engineering, The University of Hong Kong, Hong Kong (e-mails: taoyee@hku.hk, \{mycui, huangkb\}@eee.hku.hk).

Simin Wang, Shunqing Zhang, Jihong Li and Kaixuan Huang are with Shanghai Institute for Advanced Communication and Data Science, Key Laboratory of Specialty Fiber Optics and Optical Access Networks, Shanghai University, Shanghai, 200444, China (e-mails: \{siminwang, tomlijiong, xuan1999, shunqing\}@shu.edu.cn).

Wen Chen and Qingqing Wu are with the Department of Electronics Engineering, Shanghai Jiao
Tong University, Shanghai 200240, China (e-mails: \{wenchen, qingqingwu\}@sjtu.edu.cn).
}
\thanks{Corresponding author: {\em Shunqing Zhang}.}
}

\markboth{Journal of \LaTeX\ Class Files,~Vol.~14, No.~8, August~2015}%
{Shell \MakeLowercase{\textit{et al.}}: Bare Demo of IEEEtran.cls for IEEE Journals}

\maketitle

\begin{abstract}
The imminent emergence of sixth-generation (6G) networks faces critical challenges from spatially heterogeneous traffic and escalating energy consumption, necessitating sustainable scaling strategies for network infrastructure such as base stations (BSs) and reconfigurable intelligent surfaces (RISs). This paper presents a systematic scaling analysis of the Integrated Relative Energy Efficiency (IREE) metric under joint multi-BS and multi-RIS deployment in traffic-mismatched scenarios. Specifically, we propose an Alternating Directional Dual Radial Basis Function (ADD-RBF) framework that models the spatial capacity contributions of BSs and RISs through two separately parameterized RBF-type branches and maximizes IREE through accepted alternating optimization, with established representation expressiveness and stage-wise convergence properties. Theoretical analysis reveals distinct scaling behaviors: BS proliferation drives logarithmic capacity growth $\mathcal{O}(\log N^{BS})$ and polynomial large-scale mismatch reduction $\mathcal{O}\big((N^{BS})^{-t_g/2}\big)$, whereas RIS deployment provides a bounded passive capacity-gain correction and residual-structure-dependent mismatch reduction. Specifically, the RIS-side mismatch decreases polynomially as $\mathcal{O}\big((N^R)^{-t_\ell/2}\big)$ for spatially diffuse residuals and follows the stretched-exponential order $\mathcal{O}\big(\exp[-c_R\sqrt{N^R}]\big)$ for hotspot-dominated residuals. Simulation results show that RISs are effective in refining spatial traffic-capacity mismatch and alleviating hotspots, making them particularly attractive when mismatch dominates, while BSs are generally preferable under capacity shortages. These findings offer practical guidelines for green 6G network design.
\end{abstract}

\begin{IEEEkeywords}
Reconfigurable Intelligent Surfaces; Energy Efficiency; 6G Networks; Radial Basis Function; Scaling Law.
\end{IEEEkeywords}

\section{Introduction} \label{sect:intro}

The imminent arrival of sixth-generation (6G) networks promises to unlock unprecedented traffic demands fueled by immersive applications such as augmented reality, tactile internet, and remote surgery\cite{jiang2021road}. These services generate highly spatio-temporally heterogeneous traffic patterns, deviating significantly from the uniform distributions traditionally assumed in network deployment\cite{mao2021ai}. A critical consequence is the frequent emergence of localized traffic surges that overwhelm pre-provisioned network capacity, creating a pervasive mismatch between capacity and traffic distributions \cite{9888068}. Network operators often address this challenge through the dense deployment of network infrastructure. In 6G paradigms, these elements primarily encompass both Base Stations (BSs) and Reconfigurable Intelligent Surfaces (RISs) \cite{alliance2024itu}. However, this approach triggers a steep escalation in energy consumption, raising urgent concerns about the sustainability of large-scale network deployments \cite{wu2017overview}. This dichotomy compels a fundamental question: How does network performance, particularly the critical aspect of energy efficiency under realistic traffic heterogeneity, scale with the number of deployed elements, especially under a joint deployment strategy involving multiple BSs and multiple RISs? Crucially, beyond aggregate capacity scaling, understanding the scaling behavior of the root cause of inefficiency, i.e., the traffic-capacity mismatch, is paramount for guiding sustainable network evolution.

Answering this question requires more than evaluating the capacity gain of additional infrastructure. BSs and RISs improve the network in fundamentally different ways. Additional BSs introduce active transmission resources, including transmit power and bandwidth allocation, and therefore directly increase the aggregate capability. In contrast, RISs are passive elements that reshape existing BS illumination through controllable reflected paths, which can improve the spatial alignment between traffic demand and capacity supply without introducing active transmit power. Hence, joint BS/RIS deployment creates a coupled scaling tradeoff among three factors: aggregate capacity growth, traffic-capacity mismatch reduction, and additional power consumption. However, how these three effects scale with the numbers of deployed BSs and RISs, and how their interplay determines the preferred infrastructure type under different network conditions, remain insufficiently understood.

To analyze this coupled tradeoff on a common basis, this work adopts the Integrated Relative Energy Efficiency (IREE) metric, which jointly accounts for aggregate capacity, traffic-capacity mismatch, and power consumption. We develop an Alternating Directional Dual Radial Basis Function (ADD-RBF) framework for IREE-oriented joint deployment and further analyze how the scaling behaviors of these three components jointly determine the IREE improvement as the numbers of BSs and RISs increase. Our contributions are summarized as follows.

\begin{itemize}
    \item{\em Multi-BS \& Multi-RIS Enabled IREE Maximization Framework under Traffic-Capacity Mismatch.} To address the joint deployment challenge, we propose an ADD-RBF framework. This framework represents the BS direct-path and RIS-assisted spatial capacity contributions through two RBF-type dictionaries controlled by separate parameter blocks, enabling structured alternating optimization of the coupled IREE-maximization problem. Crucially, we establish the expressiveness of the ideal ADD-RBF capacity-density representation and the stage-wise local convergence property of the accepted alternating training procedure, providing theoretical support for the proposed representation and training scheme.
    \item{\em Fundamental Scaling Laws Derived from RBF Analysis.} Building upon the proposed ADD-RBF framework, we derive scaling laws for aggregate capacity and traffic-capacity mismatch in joint multi-BS/multi-RIS deployments, together with an exact marginal condition for IREE improvement. The analysis reveals distinct roles for the two types of infrastructure: BS densification provides the dominant logarithmic active-capacity growth and polynomial large-scale mismatch reduction $\mathcal{O}\big((N^{\mathrm{BS}})^{-t_g/2}\big)$, whereas RIS deployment contributes a bounded passive capacity-gain correction and refines the post-BS residual according to its spatial structure. Specifically, the RIS-side mismatch term scales as $\mathcal{O}\big((N^R)^{-t_\ell/2}\big)$ for a spatially diffuse residual and as $\mathcal{O}\big(\exp[-c_R\sqrt{N^R}]\big)$ for a hotspot-dominated residual.
    \item{\em Numerical Study and Deployment Insights.} Numerical studies illustrate the regime-dependent BS/RIS deployment tradeoff predicted by the scaling analysis. BS densification is generally favored in the capacity-limited regime, whereas RIS deployment becomes more attractive in the mismatch-limited regime. The results further indicate that aggregate traffic demand and traffic heterogeneity mainly govern the preferred BS and RIS scales, respectively.
\end{itemize}

The remainder of this paper is structured as follows. Section~\ref{sect:related} reviews related works. Section~\ref{sect:system_model} details the system model and formulates the IREE maximization problem. Section~\ref{sect:proposed_scheme} presents the ADD-RBF framework and analyzes its representation expressiveness and stage-wise convergence. Section~\ref{sect:scaling_law} analyzes the scaling behaviors of capacity and traffic-capacity mismatch, and further derives the corresponding IREE deployment insights. Section~\ref{sect:num_res} provides numerical results, and Section~\ref{sect:conc} concludes the paper.

\section{Related Work} \label{sect:related}

The scaling behavior of wireless-network performance with infrastructure density has been extensively studied through stochastic geometry, where wireless nodes are modeled as spatial point processes to characterize connectivity, interference, outage, coverage probability, and rate performance in large-scale networks \cite{haenggi2009stochastic,andrews2011tractable,elsawy2017modeling}. The homogeneous Poisson point process (PPP) provides a tractable baseline for cellular-network analysis, and has been further extended to heterogeneous cellular networks with tier-dependent densities, transmit powers, and association rules \cite{andrews2011tractable,dhillon2012modeling}. To capture more realistic spatial non-uniformity, subsequent studies have also considered non-uniform user distributions and Poisson cluster process (PCP) based heterogeneous cellular networks that account for traffic hotspots and user--small-cell correlations \cite{dhillon2013modeling,afshang2018poisson,saha2019unified}. These works establish important density-dependent statistical laws under prescribed spatial models. However, for traffic-aware green deployment, the key question is not only how average coverage or rate scales with infrastructure density, but also whether the capacity created by additional infrastructure is supplied where traffic demand is concentrated.

This spatial-alignment issue has motivated a broad range of traffic-aware green-network design studies. Large-scale user behavior and spatial traffic information have been used to guide deployment, configuration, and service control in heterogeneous cellular networks \cite{huang2014energy}. Traffic variations have also been incorporated into BS sleeping and power-control mechanisms to trade off energy saving and service delay \cite{wu2013traffic}. More recently, learning-based methods have been applied to adaptive BS activation and radio-resource management; for example, deep reinforcement learning has been used for small-cell activation in heterogeneous networks \cite{ye2020drag}, while multi-agent deep reinforcement learning has been developed for distributed dynamic power allocation in wireless networks \cite{nasir2019multi}. These studies demonstrate that incorporating traffic or network-state information can improve the efficiency of a given network operation. Nevertheless, most of them optimize conventional objectives, such as throughput, delay, power consumption, or energy efficiency, under a fixed number of infrastructure elements. As a result, they mainly answer how to optimize a given network configuration, rather than how the spatial mismatch between traffic demand and capacity supply changes as new BSs or RISs are deployed.

A mismatch-aware energy-efficiency metric is needed when the spatial capacity distribution becomes part of the performance objective. The Integrated Relative Energy Efficiency (IREE) metric addresses this issue by jointly accounting for aggregate capacity, power consumption, and traffic-capacity mismatch through the Jensen--Shannon (JS) divergence between the capacity and traffic distributions \cite{yu2022novel}. Subsequent RBF-based studies applied IREE to traffic-aware BS deployment under heterogeneous traffic \cite{yu2024iree,yu2025model}. However, these studies focus primarily on algorithmic optimization with a fixed number of BSs, without characterizing the scaling behavior of IREE components or the mismatch-reduction role of passive RIS deployment.

Reconfigurable intelligent surfaces have been widely studied as a promising technique for constructing smart radio environments through programmable passive reflections \cite{pan2021reconfigurable,9140329}. Representative studies have investigated joint active/passive beamforming \cite{wu2019intelligent}, channel estimation and hardware-constrained passive beamforming \cite{zheng2022survey,wu2021intelligent}, as well as RIS-assisted energy-efficient resource allocation and distributed RIS control \cite{huang2019reconfigurable,yang2021energy}. These studies demonstrate the potential of RISs to improve link quality and energy efficiency through passive signal reconfiguration. In parallel, RIS-related physical scaling behaviors have been analyzed from the perspectives of path-loss modeling, aperture size, near-field propagation, and power scaling laws \cite{tang2020wireless,bjornson2020power}. However, these works mainly focus on link-level beamforming/resource allocation or surface-level physical scaling. They do not characterize how increasing the number of deployed RIS panels reshapes the network-level spatial capacity density, nor how this passive mismatch-correction capability should be balanced with active BS densification and infrastructure power consumption under the IREE metric.

In summary, existing studies provide important but separate insights into stochastic density scaling, traffic-aware fixed-size optimization, mismatch-aware energy-efficiency evaluation, and RIS-assisted transmission. What remains missing is a unified scaling framework for optimized joint BS/RIS deployment that jointly characterizes aggregate capacity growth, traffic-capacity mismatch reduction, and infrastructure power consumption. This paper addresses this gap by developing an IREE-oriented ADD-RBF framework, analyzing the capacity and mismatch scaling behaviors, and clarifying the distinct roles of BSs and RISs in capacity-limited and mismatch-limited regimes.


\section{System Model and Problem Formulation} 
\label{sect:system_model}

In this section, we present a multi-BS/multi-RIS MISO system and introduce an IREE-maximization problem for joint deployment to optimize the active beamforming vectors, the bandwidths and the locations of BSs, as well as the passive beamforming matrices and the locations of RISs.

\subsection{System Model}
\label{subsec:sys}

\begin{figure}[t] 
\centering  
\includegraphics[height=4.5cm,width=8cm]{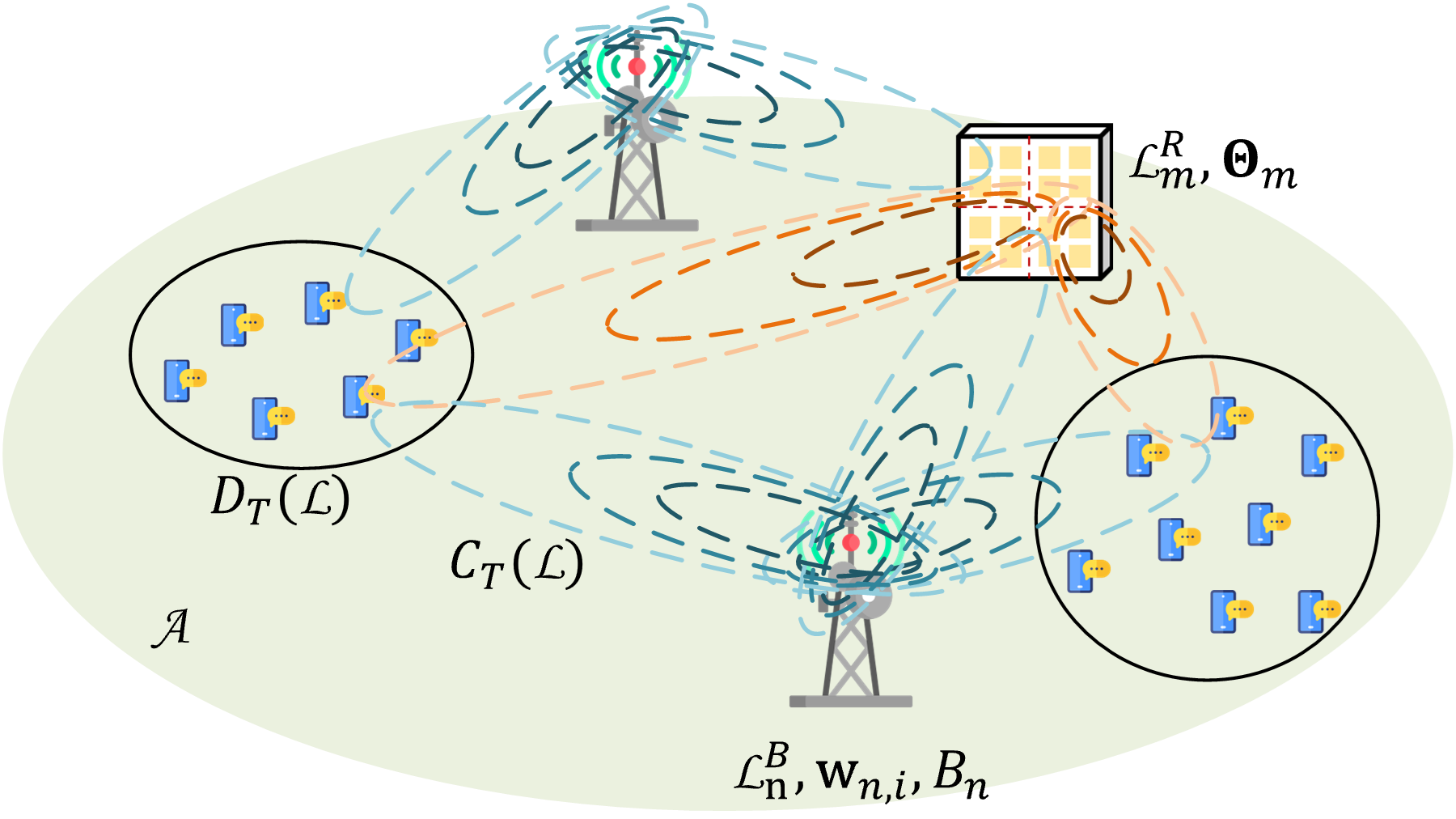}
\caption{An illustrative example of target area $\mathcal{A}$ with $N^{BS}$ BSs and $N^R$ RISs. The active beamforming vectors, the bandwidths and the locations of BSs, as well as the passive beamforming matrices and the locations of RISs shall be optimized to maximize the energy efficiency of the network.}
\label{fig:scenario}
\end{figure}

Consider a frequency division wireless network consisting of $N^{BS}$ BSs with $N^{BS}_{T}$ antennas and $N^{R}$ RISs with $N^{R}_{T}$ elements as shown in Fig.~\ref{fig:scenario}. Denote $\mathbf{H}^{br}_{n,m} \in \mathbb{C}^{N^{BS}_{T} \times N^{R}_{T}}$, $\mathbf{h}^{bu}_{n} \in \mathbb{C}^{N^{BS}_{T} \times 1} $, and $\mathbf{h}^{ru}_{m} \in \mathbb{C}^{N^{R}_{T} \times 1}$ as the channel coefficients between the $n$-th BS and the $m$-th RIS, the $n$-th BS and the user, and the $m$-th RIS and the user, respectively. With the passive beamforming matrix at the $m$-th RIS, $\mathbf{\Theta}_{m} \in  \mathbb{C}^{N^{R}_{T} \times N^{R}_{T}}$, the hybrid channel condition between the $n$-th BS and the user is given by
\begin{eqnarray}
\label{eqn:channels}
    \mathbf{h}_{n} & = & \sum^{N^{R}}_{m=1} \mathbf{H}^{br}_{n,m} \mathbf{\Theta}_{m} \mathbf{h}^{ru}_{m} + \mathbf{h}^{bu}_n,
\end{eqnarray}

Let $\mathbf{w}_{n,i}$ denote the $i$-th beamforming vector of the $n$-th BS. The relationship between the transmit signal $s_{n,i}$ and the received signal $y_n$ is then given by
\begin{eqnarray}
y_{n} & = & \mathbf{h}_{n}^H \sum_{i=1}^{N^{BS}_{T}} \mathbf{w}_{n,i} s_{n,i} + z_{n}, 
\end{eqnarray}
where $z_{n}$ represents the additive white Gaussian noise (AWGN) with zero mean. The received SNR at the user side is therefore given by
\begin{eqnarray} \label{eq:capacity}
r_{n} & = & \frac{\sum_{i=1}^{N^{BS}_{T}} \left \Vert \mathbf{h}_{n}^H  \mathbf{w}_{n,i} \right \Vert_2^2}{\sigma^2 B_{n}},
\end{eqnarray}

where $B_n$ denotes the effective bandwidth resource assigned to the $n$-th BS, and $\sigma^2$ represents the power spectral density of $z_n$. The aggregate bandwidth allocation satisfies
$\sum_{n=1}^{N^{BS}}B_n\leq B_{\max}$, where $B_{\max}$ denotes the aggregate bandwidth budget over the considered service area.\footnote{In this work, we focus on network-level deployment rather than detailed carrier assignment and frequency-reuse design. Under this abstraction, $B_n$ denotes the effective bandwidth available to the $n$-th BS under the adopted spectrum allocation and coordination scheme. Accordingly, $B_{\max}$ represents the aggregate bandwidth budget over the considered service area.
}
By summing over all $N^{BS}$ BSs, the spatial capacity density at location $\mathcal{L}$ is given by
\begin{eqnarray} \label{eqn:tot_cap}
C_T(\mathcal{L}) &=& \frac{1}{|\mathcal A|} \sum_{n=1}^{N^{BS}}  B_{n} \log_2 (1 + r_{n}),
\end{eqnarray}
where $|\mathcal A|$ denotes the area of the considered service region.

The total power consumption of the $n$-th BS is given by $ P_n = \lambda \sum_{i=1}^{N_T^{BS}} \|\mathbf w_{n,i}\|_2^2 + P^c$, where $\lambda$ denotes the amplification coefficient associated with the power-amplifier efficiency \cite{xu2012energy}, and $P^c$ denotes the static circuit power of a BS. Therefore, the total amount of power consumption $P_T$ is given by 
\begin{eqnarray}
\label{eqn:total_pow_def}
P_T = \lambda \sum_{n=1}^{N^{BS}} \sum^{N^{BS}_{T}}_{i=1} \|\mathbf{w}_{n,i}\|_2^2 + N^{R}P^{r} +  N^{BS} P^{c},
\end{eqnarray}
where $P^{r}$ denotes the static circuit power of the RIS.

The following assumptions are adopted throughout the rest of this paper.
First, the spatial traffic density over the target area $\mathcal{A}$ is denoted by $D_T(\mathcal L)$, and it is assumed to be continuous over $\mathcal A$. Second, let $\mathcal L^U$, $\mathcal L_n^B$, and $\mathcal L_m^R$
denote the locations of the user, the $n$-th BS, and the
$m$-th RIS, respectively. The wireless channel is given by
\begin{equation}
\label{eqn:channel_model}
\left\{ 
\begin{aligned}
\mathbf{h}^{bu}_n &= \mathbf{a}^{bu}/L(\mathcal{L}^U, \mathcal{L}^B_n) \\
\mathbf{h}^{ru}_m &= \mathbf{a}^{ru}/L(\mathcal{L}^U, \mathcal{L}^R_{m}) \\
\mathbf{H}^{br}_{n,m} &= \mathbf{a}^{br} (\mathbf{a}^{ru})^H/ L(\mathcal{L}^R_{m}, \mathcal{L}^B_n) 
\end{aligned}
\right .
\end{equation}
where $\mathbf{a}^{bu}, \mathbf{a}^{br} \in \mathbb{C}^{N^{BS}_{T} \times 1}$ and $\mathbf{a}^{ru} \in \mathbb{C}^{N^{R}_{T} \times 1}$ are the steering
vectors of the BS and the RIS, respectively. $L(\mathcal{L}, \mathcal{L}')$ denotes the attenuation coefficient related to path loss given by $L(\mathcal{L}, \mathcal{L}') = \gamma \Vert \mathcal{L} - \mathcal{L}'\Vert^{\alpha} + \beta$, where $\alpha$ is the path-loss exponent, while $\gamma>0$
and $\beta\geq0$ are the corresponding path-loss coefficients. During the evaluation period, $\beta, \gamma, \lambda$, $P^{c}$ and $P^{r}$ are assumed to be constant\footnote{The non-constant channel fading effects, such as shadowing, will be discussed through numerical results in Section~\ref{sect:num_res}.}. Third, the phase-shift matrix is assumed to be diagonal, i.e.,
$\mathbf{\Theta}_{m}
=
\operatorname{diag}\{\boldsymbol{\theta}_m\}
\in
\mathbb{C}^{N^R_T \times N^R_T}$
for any RIS $m$, where 
$\boldsymbol{\theta}_m
=
[\theta_{m,1},\theta_{m,2},\ldots,\theta_{m,N^R_T}]^T
\in
\mathbb{C}^{N^R_T \times 1}$.
Each reflecting coefficient is parameterized as
$
\theta_{m,j}=e^{\mathrm{i}\phi_{m,j}}, j \in \{1,2,\ldots,N^R_T\}$ with $\phi_{m,j}\in[0,2\pi)$ denoting the continuous phase shift of the $j$-th reflecting element of the $m$-th RIS. Last but not least, we assume that each RIS can reflect signals from all BSs in $\mathcal{A}$\footnote{In fact, the framework proposed in this paper can also be applied to the scenario where each RIS can only reflect the signals from its local BS.}.

\subsection{Problem Formulation}
\label{subsec:formu}

Conventional Energy Efficiency (EE) defined by the ratio of total throughput to total energy consumption fails to capture the traffic-capacity mismatch, and therefore overlooks the fact that high aggregate capacity may still lead to poor performance if the capacity is poorly distributed \cite{mirahsan2015hethetnets}. To simultaneously consider the absolute capacity provision and spatial matching, in this paper we adopt the IREE metric which incorporates the JS divergence to explicitly quantify traffic-capacity mismatch. Specifically, we can define the IREE by incorporating the traffic-capacity mismatch as follows.

\begin{Def}[IREE Metric \cite{yu2022novel}] \label{def:IREE}
The IREE of wireless networks, $\eta_{IREE}$, is defined as,
\begin{eqnarray} \label{eq:def_iree}
\eta_{IREE} = \frac{\min\{C_{Tot},D_{Tot}\}\left[1 - \xi\left(C_{T}, D_{T}\right)\right]}{P_T}.
\end{eqnarray}
In the above expression, $C_{Tot} = \iint_{\mathcal{A}}C_{T}(\mathcal{L}) \textrm{d}\mathcal{L}$ and $D_{Tot} = \iint_{\mathcal{A}}D_{T}(\mathcal{L}) \textrm{d}\mathcal{L}$ denote the aggregate capacity and the aggregate traffic demand over the whole evaluation area $\mathcal{A}$. $\xi\left(C_{T}, D_{T}\right)$ is the JS divergence given by $\frac{1}{2} \iint_{\mathcal{A}}  \frac{C_{T}(\mathcal{L})}{C_{Tot}} \log_2 \left[ \frac{2 D_{Tot} C_{T}(\mathcal{L}) }{ D_{Tot} C_{T}(\mathcal{L}) +  C_{Tot} D_{T}(\mathcal{L}) } \right] + \frac{D_{T}(\mathcal{L})}{D_{Tot}} \log_2 \left[ \frac{2  C_{Tot} D_{T}(\mathcal{L}) }{ C_{Tot} D_{T}(\mathcal{L}) + D_{Tot} C_{T}(\mathcal{L}) } \right] \textrm{d}\mathcal{L}$.
\end{Def}

Consequently, the IREE metric jointly accounts for aggregate capacity and traffic-capacity mismatch. The corresponding maximization problem is formulated as follows.

\begin{Prob}[Original IREE-Maximization Problem for Joint Deployment] \label{prob:origin}
The IREE of multi-BS/multi-RIS MISO system in Section~\ref{subsec:sys} can be maximized by the following joint deployment problem.
\begin{eqnarray}
    \underset{\makecell{\{\mathcal{L}^B_n\}, \{\mathbf{w}_{n,i}\}, \{B_n\} \\ \{\mathcal{L}^R_{m}\}, \{\mathbf{\Theta}_{m}\} }}{\textrm{maximize}} && \eta_{IREE}, \nonumber \\
    \textrm{subject to} 
    && \zeta(\{\mathcal{L}^B_n\}, \{\mathbf{w}_{n,i}\}, \{B_n\},\nonumber \\ 
    && \{\mathcal{L}^R_{m}\}, \{\mathbf{\Theta}_{m}\}) \geq \zeta_{\min}, \label{constrain:qos} \\
    && \sum_{n=1}^{N^{BS}} B_n \leq B_{\max},  B_n > 0, \forall n, \label{constrain:bandwidth} \\
    && \ \sum^{N^{BS}_{T}}_{i=1} \|\mathbf{w}_{n,i}\|_2^2 \leq P_{\max}, \forall n. \label{constrain:max_power} 
\end{eqnarray}
In the above optimization problem, $B_{\max}$ is the aggregate bandwidth budget shared by all BSs and $P_{\max}$ is the power limit for a single BS. 
Define
$\zeta(\{\mathcal{L}^B_n\},\{\mathbf{w}_{n,i}\},\{B_n\},
\{\mathcal{L}^R_m\},\{\mathbf{\Theta}_m\})
=
\frac{\min\{C_{Tot},D_{Tot}\}
[1-\xi(C_T,D_T)]}{D_{Tot}}$
as the customer satisfaction score (CSAT).
Constraint~\eqref{constrain:qos} guarantees a minimum CSAT level
$\zeta_{\min}\in[0,1]$.
\end{Prob}

Problem~\ref{prob:origin} is typically a fractional programming problem, which, according to Dinkelbach's algorithm \cite{dinkelbach1967nonlinear}, can be efficiently solved by the following two iterative steps. First, solve the utility maximization problem defined in Problem~\ref{prob:transformed} for a given $\eta_{IREE}^{(k)}$. Then, use the obtained network parameters $\{\mathcal{L}^{B,(k)}_n\}, \{\mathbf{w}^{(k)}_{n,i}\}, \{B^{(k)}_n\}, \{\mathcal{L}^{R,(k)}_{m}\}, \{\mathbf{\Theta}^{(k)}_{m}\}$ to update $\eta_{IREE}^{(k+1)}$. This two-step process repeats until convergence.

\begin{Prob}[Utility-Maximization Problem for a Given IREE] 
\label{prob:transformed}
For any given IREE, the utility function, $\min\{C_{Tot},D_{Tot}\}\big[1 - \xi(C_{T}, D_{T}) \big] - \eta_{IREE} P_T$, can be maximized via the following optimization problem.
\begin{eqnarray}
    \underset{\makecell{\{\mathcal{L}^B_n\}, \{\mathbf{w}_{n,i}\}, \{B_n\} \\ \{\mathcal{L}^R_{m}\}, \{\mathbf{\Theta}_{m}\} }}{\textrm{maximize}} && \min\{C_{Tot},D_{Tot}\}\left[1 - \xi\left(C_{T}, D_{T}\right)\right] \nonumber \\
    && -\eta_{IREE} P_T, \\
    \textrm{subject to} &&  \eqref{constrain:qos} - \eqref{constrain:max_power}.  \nonumber 
\end{eqnarray}
\end{Prob}

However, Problem~\ref{prob:transformed} remains challenging due to three inherent difficulties. First, the severe non-convexity of the JS divergence metric $\xi(C_T, D_T)$ is exacerbated by RIS-induced cascaded channels, which intensifies variable coupling and invalidates conventional convex optimization. Second, the objective requires the capacity distribution to match a heterogeneous traffic distribution over the entire continuous service region, rather than optimizing a finite set of link-level performance variables. This spatial-field matching requirement substantially increases the dimensionality and computational complexity of the joint deployment problem. Third, conventional single-branch RBF methods \cite{yu2024iree} cannot be directly applied because the RIS-assisted spatial contribution is jointly determined by the BS--RIS and RIS--user propagation paths and the controllable RIS phases, rather than by a standard distance-based kernel with an independently adjustable center.

\section{Proposed ADD-RBF Scheme} 
\label{sect:proposed_scheme}

In this section, we develop the proposed ADD-RBF scheme for IREE-oriented joint BS/RIS deployment. The LoS-related BS direct-path and RIS-assisted reflected components are modeled as two coupled RBF branches and optimized through accepted alternating updates. We then analyze the representation expressiveness and stage-wise convergence of the resulting scheme.

\subsection{Dual-RBF Architecture}

\begin{figure*}[t]
\centering  
\includegraphics[height=7cm,width=17cm]{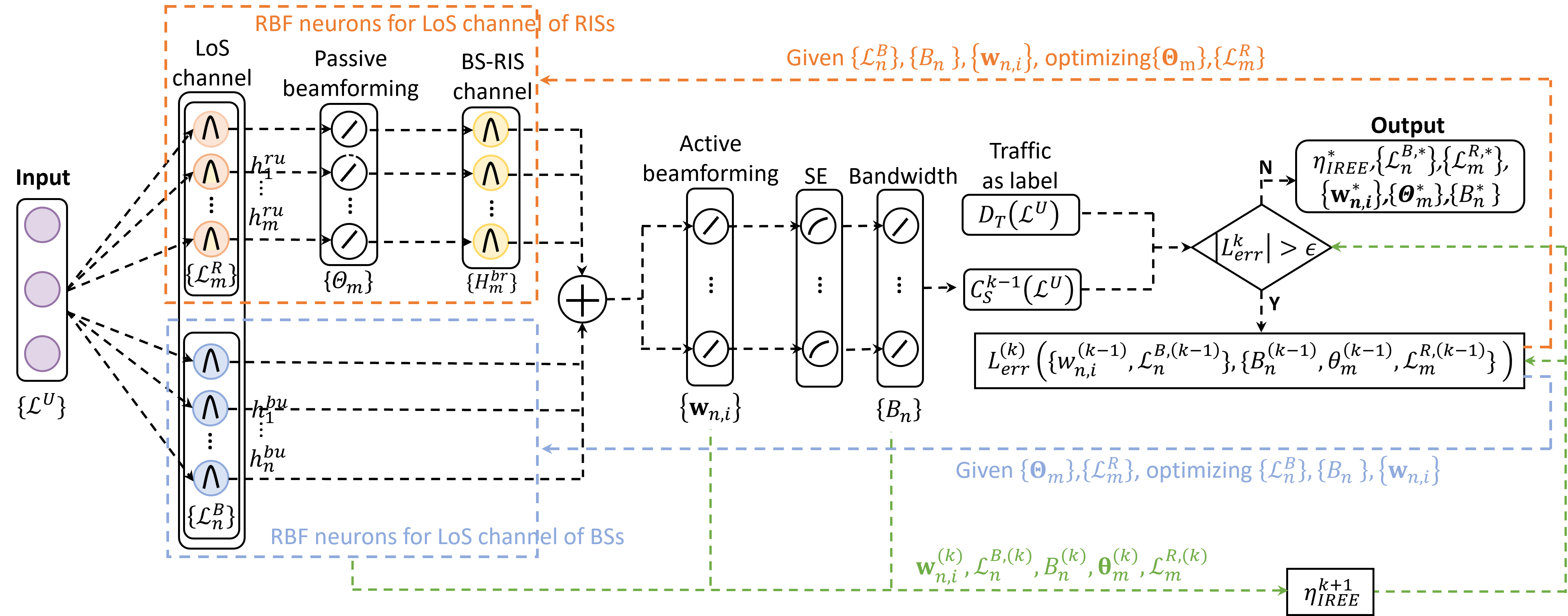}
\caption{Overview of the proposed ADD-RBF scheme. The black dashed line represents the forward propagation while the orange and blue dashed lines represent the backward propagation processes for the LoS channel of the RISs and BSs, respectively. The green dashed line illustrates how the optimized IREE is obtained} through a series of $L_{err}^{(k)}$ minimization problems, where $L_{err}^{(k)}$ is constructed through IREE in current iteration $\eta_{IREE}^{(k)}$.
\label{fig:algorithm}
\end{figure*}

Based on the spatial capacity density defined in \eqref{eqn:tot_cap},
a lower bound $C_S(\mathcal L^U)$ on $C_T(\mathcal L)$ is given by
\begin{eqnarray}
\label{eqn:def_CsB} 
C_T(\mathcal{L}) \geq \frac{1}{|\mathcal A|} \sum_{n=1}^{N^{BS}} B_n S_n(\mathcal{L}^U, \mathcal{L}^B_n, \{ \mathcal{L}^R_{m} \}) \triangleq C_S(\mathcal{L}^U), 
\end{eqnarray}
where $S_n(\mathcal{L}^U, \mathcal{L}^B_n, \{\mathcal{L}^R_{m}\}) = \log_2 \left( 1 + \frac{\sum_{i=1}^{N^{BS}_{T}} \left \Vert \mathbf{h}_{n}^H  \mathbf{w}_{n,i} \right \Vert_2^2}{\sigma^2 B_{\max} } \right)$.

Radial basis functions are real-valued functions whose value depends solely on the distance from a fixed point (the center). Under the above mathematical transformation, we have the following lemma.

\begin{Lem}[Dual-RBF Architecture]
\label{lem:dual-rbf}
The spatial capacity representation $C_S(\mathcal L^U)$ admits a dual-dictionary structure composed of a BS direct-path RBF branch and an RIS-assisted reflected RBF-type branch. The BS branch is formed by radial path-loss atoms centered at the BS locations, whereas each RIS-assisted atom combines a radial RIS--user path-loss envelope with a phase-controlled array factor. The two branches are controlled by separate BS-side and RIS-side parameter blocks and can therefore be optimized alternately.
\end{Lem}
\IEEEproof Please refer to Appendix~\ref{appendix:dual-rbf} for the proof.
\endIEEEproof

As illustrated in Fig.~\ref{fig:algorithm}, the BS and RIS channel components induce different RBF-type spatial structures, which motivates the alternating optimization of BS and RIS configurations. Lemma~\ref{lem:dual-rbf} further shows that $C_S(\mathcal L^U)$ admits a dual-dictionary representation composed of a BS direct-path RBF dictionary and an RIS-assisted reflected dictionary. This dual-dictionary viewpoint allows us to characterize the approximation capability of the proposed architecture.

\begin{Thm}[Expressiveness of the Dual-RBF Capacity Density Representation]
\label{thm:Existence_risRBF}
Let $\mathcal A\subset\mathbb R^2$ be a compact service region, and let $D_T(\mathcal L)$ be any bounded continuous non-negative traffic distribution over $\mathcal A$. 
Consider the ideal dual-RBF capacity density family induced by $C_S(\mathcal L)$ in \eqref{eqn:def_CsB}, where the number of RBF atoms and their representation parameters are allowed to be sufficiently rich. Then, for any $\epsilon>0$, there exists a set of ADD-RBF representation parameters such that $\sup_{\mathcal L\in\mathcal A}
\left|
C_S(\mathcal L)-D_T(\mathcal L)
\right|
<\epsilon$.
\end{Thm}

\IEEEproof
Let $\mathcal F_{\rm ADD}$ denote the dual-RBF capacity density family and $\mathcal F_{\rm B}$ denote its BS direct-path subfamily. By Lemma~\ref{lem:dual-rbf}, the RIS-assisted dictionary enlarges the representation without removing the BS direct-path dictionary, hence $\mathcal F_{\rm B}\subseteq \mathcal F_{\rm ADD}$. Since the BS-RBF family $\mathcal F_{\rm B}$ is dense in the space of bounded continuous non-negative functions over compact $\mathcal A$ according to the classical RBF approximation result adopted in \cite{yu2024iree}, for any $\epsilon>0$ there exists $C_B\in\mathcal F_{\rm B}$ such that $\sup_{\mathcal L\in\mathcal A}|C_B(\mathcal L)-D_T(\mathcal L)|<\epsilon$. The inclusion $\mathcal F_{\rm B}\subseteq \mathcal F_{\rm ADD}$ gives the desired dual-RBF representation. 
\endIEEEproof

Theorem~\ref{thm:Existence_risRBF} establishes the representation foundation of the proposed dual-RBF architecture by showing that the addition of the RIS-assisted dictionary preserves the universal approximation capability inherited from the BS-RBF subfamily. This result characterizes the expressiveness of the ideal ADD-RBF family. For a practical network with finite numbers of BSs and RISs, the achievable traffic-capacity mismatch further depends on the deployment scale and the spatial structure of the remaining mismatch, as characterized in Section~\ref{sect:scaling_law}.

\subsection{Alternating Training Scheme}

The dual-RBF architecture in Lemma~\ref{lem:dual-rbf} provides a structured capacity density representation for the joint BS/RIS deployment problem. We next describe how this representation is trained to solve the Dinkelbach-transformed problem in Problem~\ref{prob:transformed}. Since the objective contains the spatial JS divergence term and the cascaded BS-RIS-user channel, the BS-side variables and RIS-side variables are strongly coupled. Directly updating all variables in an end-to-end manner may therefore lead to unstable training trajectories. To address this issue, we develop an accepted alternating training scheme, in which the BS-side and RIS-side variable blocks are updated sequentially while the other block is fixed.

For a given Dinkelbach parameter $\eta_{IREE}^{(k)}$, let
$\mathcal X_B=\{\mathcal L_n^B,B_n,\mathbf w_{n,i}\}$ and
$\mathcal X_R=\{\mathcal L_m^R,\boldsymbol{\theta}_m\}$
denote the BS-side and RIS-side variable blocks, respectively, and define
$\mathcal X_{all}=(\mathcal X_B,\mathcal X_R)$. We construct a sampled training loss from the negative transformed utility in Problem~\ref{prob:transformed}. Specifically, over $Q$ uniformly sampled evaluation points $\{\mathcal L_q^U\}_{q=1}^{Q}$, the loss is defined as
\begin{small}
\begin{eqnarray}
&L_{err}(\mathcal X_{all};\kappa)
=-\frac{|\mathcal A|}{Q}
\min\left\{\sum_{q=1}^{Q}C_S(\mathcal L_q^U),
\sum_{q=1}^{Q}D_T(\mathcal L_q^U)\right\}
\nonumber\\
&\times\left[1-\xi_Q(C_S,D_T)\right]
+\eta_{IREE}^{(k)}P_T+\kappa\Omega(\mathcal X_{all}).
\label{eqn:loss_func}
\end{eqnarray}
\end{small}
where $\xi_Q \left(C_S, D_{T} \right)$ is the sample-based approximation of the JS divergence obtained by replacing the spatial integrals with uniform quadrature over $\{\mathcal L_q^U\}_{q=1}^{Q}$. The corresponding penalty term is given by $\Omega(\mathcal X_{all}) = \max \big\{ \zeta_{\min}  - \zeta(\mathcal X_{all}), 0  \big\}  + \sum_{n=1}^{N^{BS}} \max \big \{ \sum^{N^{BS}_{T}}_{i=1} \|\mathbf{w}_{n,i}\|_2^2 - P_{\max}, 0  \big\} + \max \left \{ \sum_{n=1}^{N^{BS}} B_n - B_{\max}, 0  \right \}$ with $\kappa$ the penalty coefficient. Thus, minimizing $L_{err}$ improves the sampled surrogate of the transformed utility while penalizing constraint violations.

In each training round, we first update $\mathcal X_B$ with $\mathcal X_R$ fixed, and then update $\mathcal X_R$ with the updated $\mathcal X_B$ fixed. For each block update, Adam is used to generate a candidate solution. The candidate is accepted only when it satisfies the sufficient-decrease condition
\begin{eqnarray}
L_{err}(\widetilde{\mathcal X};\kappa)
\leq
L_{err}(\mathcal X;\kappa)-
\tau
\left\|
\widetilde{\mathcal X}-\mathcal X
\right\|^2,
\label{eqn:sufficient_decrease}
\end{eqnarray}
where $\mathcal X$ denotes the current BS-side or RIS-side block, $\widetilde{\mathcal X}$ denotes the corresponding Adam-generated candidate, and $\tau>0$ is a small constant. If \eqref{eqn:sufficient_decrease} is not satisfied, the learning rate is reduced by a factor $\rho_\mu\in(0,1)$ and a new candidate is generated. If no candidate is accepted after $N_{bt}$ backtracking trials, the current block is kept unchanged. During each candidate evaluation, all location-dependent channels are recomputed according to \eqref{eqn:channel_model}. In particular, $\mathbf H_{n,m}^{br}$ is updated whenever either $\mathcal L_n^B$ or $\mathcal L_m^R$ changes.

The training is performed in two sequential stages. In the first stage, we set $\kappa=0$ and minimize the unconstrained transformed loss to obtain a warm start. In the second stage, we set $\kappa=\kappa_{large}$ and continue the accepted alternating training to refine feasibility with respect to the CSAT, bandwidth, and power constraints. The sufficient-decrease rule in \eqref{eqn:sufficient_decrease} is applied separately within each fixed-$\kappa$ stage. To ensure well-defined SNR and RIS phase expressions, $B_n$ is parameterized as a positive variable and each RIS coefficient is parameterized as $\theta_{m,j}=e^{\mathrm{i} \phi_{m,j}}$. The remaining constraints are handled through the penalty term in \eqref{eqn:loss_func}.

For each fixed penalty coefficient $\kappa$ and Dinkelbach-type parameter $\eta_{IREE}^{(k)}$, the accepted BS-side and RIS-side updates are repeated until the inner stopping criterion is satisfied or the maximum number of inner iterations is reached. The resulting deployment variables are then substituted into the original IREE definition in \eqref{eq:def_iree} to update $\eta_{IREE}^{(k+1)}$ for the next transformed loss. Since the resulting subproblems are non-convex and are solved by inexact
local updates, the instantaneous IREE is not required to increase after every inner block update. Instead, the algorithm records the feasible solution with the largest IREE among the solutions obtained after the inner alternating loops throughout the two-stage training and returns it as the final deployment output.
The complete procedure is summarized in Algorithm~\ref{alg:Dinkelbach}.

\IncMargin{1em}
\begin{algorithm}[t]
\caption{Proposed ADD-RBF Scheme}
\label{alg:Dinkelbach}
\SetKwInOut{Input}{input}
\SetKwInOut{Output}{output}
\SetKwFunction{AccUp}{AccUp}
\SetKwProg{Fn}{Function}{:}{}

\Input{Traffic density $D_T(\mathcal L)$; system parameters in
Table~\ref{tab:simu_para}; and training parameters
$\{N_B^{Adam}$, $N_R^{Adam}$, $\mu_B$, $\mu_R$, $\rho_\mu$, $N_{bt}$, $\kappa_{large}$, $\tau$, $\epsilon_{in}$, $N_{in}^{\max}$, $\epsilon_{stop}\}$.}

Initialize
$\mathcal X_B=\{\mathcal L_n^B,B_n,\mathbf w_{n,i}\}$,
$\mathcal X_R=\{\mathcal L_m^R,\boldsymbol{\theta}_m\}$, and
$\mathcal X=(\mathcal X_B,\mathcal X_R)$; parameterize
$B_n>0$ and $\theta_{m,j}=e^{\mathrm{i}\phi_{m,j}}$;

Set $\eta_{IREE}^{best}=-\infty$ and initialize the recorded
feasible solution as empty;

\Fn{\AccUp{$\mathcal X,\mathcal Y,\mu,N_{Adam},\kappa$}}{
$\mu_{bt}\leftarrow\mu$;

\For{$r=1$ \KwTo $N_{bt}$}{
Generate $\widetilde{\mathcal X}$ from the current $\mathcal X$
using $N_{Adam}$ Adam steps on
$L_{err}(\mathcal X,\mathcal Y;\kappa)$ with learning rate
$\mu_{bt}$ and fixed $\mathcal Y$;

Recompute the affected location-dependent channels;

\If{
$L_{err}(\widetilde{\mathcal X},\mathcal Y;\kappa)
\leq
L_{err}(\mathcal X,\mathcal Y;\kappa)
-\tau\|\widetilde{\mathcal X}-\mathcal X\|^2$
}{
\textbf{return} $\widetilde{\mathcal X}$;
}

$\mu_{bt}\leftarrow\rho_\mu\mu_{bt}$;
}

\textbf{return} $\mathcal X$;
}

\tcp{Sequential two-stage training}

\For{$\kappa\in\{0,\kappa_{large}\}$}{

Reset the Adam learning rates and moment estimates; compute
$\eta_{old}$ from the current $\mathcal X$ and set
$\Delta_\eta=+\infty$;

\While{$\Delta_\eta>\epsilon_{stop}$}{

Construct $L_{err}(\cdot;\kappa)$ using fixed $\eta_{old}$;
set $\Delta_{in}=+\infty$ and $s=0$;

\While{
$\Delta_{in}>\epsilon_{in}
\ \mathrm{and}\
s<N_{in}^{\max}$
}{

$\mathcal X^{old}\leftarrow\mathcal X$;

$\mathcal X_B\leftarrow$
\AccUp{$\mathcal X_B,\mathcal X_R,\mu_B,
N_B^{Adam},\kappa$};

$\mathcal X_R\leftarrow$
\AccUp{$\mathcal X_R,\mathcal X_B,\mu_R,
N_R^{Adam},\kappa$};

Update $\mathcal X=(\mathcal X_B,\mathcal X_R)$,
$\Delta_{in}\leftarrow
\|\mathcal X-\mathcal X^{old}\|^2$, and $s\leftarrow s+1$;
}

Compute $\eta_{new}$ from \eqref{eq:def_iree};

\If{
the current solution is feasible and
$\eta_{new}>\eta_{IREE}^{best}$
}{
$\eta_{IREE}^{best}\leftarrow\eta_{new}$ and record
the current $\mathcal X$;
}

$\Delta_\eta\leftarrow|\eta_{new}-\eta_{old}|$ and
$\eta_{old}\leftarrow\eta_{new}$;
}
}

\Output{The recorded best feasible $\eta^{\star}_{IREE}$, $\{\mathcal{L}_{n}^{B,\star}\}$, $\{\mathcal{L}_{m}^{R,\star}\}$, $\{\mathbf{w}_{n,i}^{\star}\}$, $\{\mathbf{\Theta}_{m}^{\star}\}$ and $\{B_{n}^{\star}\}$.}

\end{algorithm}
\DecMargin{1em}

\subsection{Performance Analysis}

We next discuss the training behavior of the proposed ADD-RBF scheme. Since the BS-side and RIS-side variables are coupled through the cascaded channel and the JS divergence-based mismatch term, the training trajectory may oscillate when all variables are updated simultaneously. The accepted alternating update in Algorithm~\ref{alg:Dinkelbach} is used to stabilize this process. Since the two-stage training sequentially uses $\kappa=0$ and $\kappa=\kappa_{large}$, the following result characterizes the limiting behavior of the accepted inner alternating sequence for each fixed pair of $\kappa$ and $\eta_{IREE}^{(k)}$. The parameters $\epsilon_{in}$ and $N_{in}^{\max}$ provide practical finite stopping controls for this inner procedure.

\begin{Thm}[Stage-wise Convergence of ADD-RBF Training]
\label{thm:conver_analy}
For each fixed penalty stage $\kappa\in \{0,\kappa_{large}\}$ and each fixed Dinkelbach-type parameter $\eta_{IREE}^{(k)}$ within this stage, consider the accepted alternating BS-side and RIS-side sequence
underlying the inner loop of Algorithm~\ref{alg:Dinkelbach}. Assume that the accepted variable sequence is bounded, $L_{err}(\cdot;\kappa)$ is lower bounded, and the accepted Adam updates are first-order consistent\footnote{Here, first-order consistency means that, after backtracking, the accepted Adam-generated candidate can realize a sufficiently small descent step whenever a local block descent direction exists.}. Then, within the corresponding fixed-$\kappa$ and fixed-$\eta_{IREE}^{(k)}$ training round, the accepted loss sequence is non-increasing and converges to a finite value. Moreover, the difference between two consecutive accepted updates vanishes asymptotically. Every accumulation point of the accepted sequence is a block-stationary point of $L_{err}(\cdot;\kappa)$.
\end{Thm}
\IEEEproof
Please refer to Appendix~\ref{appendix:conver_analy}.
\endIEEEproof

Theorem~\ref{thm:conver_analy} establishes the local convergence property of the accepted inner alternating sequence for each fixed pair of $\kappa$ and $\eta_{IREE}^{(k)}$. Upon practical termination of the inner loop, the Dinkelbach-type parameter is updated and the next transformed loss is constructed. Since these transformed subproblems are solved by inexact local updates, Algorithm~\ref{alg:Dinkelbach} does not require the instantaneous IREE to increase after every inner block update. Instead, it records the best feasible IREE among the solutions obtained after the inner alternating loops and uses the corresponding deployment as the final output.

\begin{figure}[t]
\centering  
\includegraphics[height=6cm,width=7.5cm]{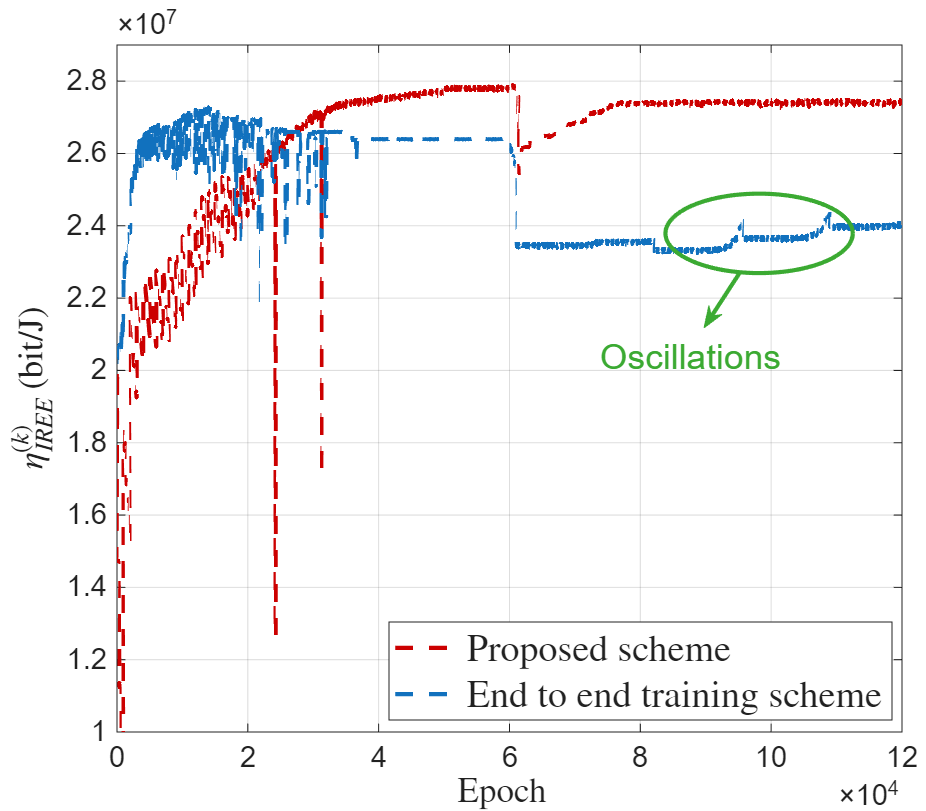}
\caption{Training comparison between the proposed ADD-RBF scheme and the end-to-end training baseline. The proposed scheme exhibits more stable training behavior due to the accepted alternating updates, while the end-to-end baseline shows stronger oscillations caused by the coupled BS/RIS variables.}
\label{fig:training_compare}
\end{figure}

To validate the practical training behavior, we compare the proposed ADD-RBF training with the end-to-end training baseline in Fig.~\ref{fig:training_compare}. The proposed scheme exhibits more stable training behavior due to the accepted alternating updates, while the end-to-end baseline exhibits stronger oscillations caused by the simultaneous update of coupled BS and RIS variables.

\section{IREE Scaling Analysis for Joint Deployment}
\label{sect:scaling_law}

In this section, we characterize how capacity, power consumption, and traffic-capacity mismatch jointly determine the IREE performance under increasing numbers of BSs and RISs. We first derive the capacity scaling and the exact marginal-gain condition for additional infrastructure. We then analyze the mismatch scaling induced by BS direct-path and RIS-assisted reflected atoms. Finally, we translate these results into deployment guidelines for the capacity-limited and mismatch-limited regimes.

\subsection{Capacity Scaling and IREE Gain Condition}
\label{subsec:cap_power_gain}

According to Definition~\ref{def:IREE}, IREE is jointly determined by the capacity term $\min\{C_{Tot},D_{Tot}\}$, the mismatch term $\xi(C_T,D_T)$, and the power term $P_T$.

Under the per-BS power constraint, Appendix~\ref{appendix:capacity_scaling} gives the order-wise form of the aggregate capacity
\begin{small}
\begin{equation}
C_{Tot}
=
\mathcal O
\Bigg(
B_{\max}
\log_2
\Bigg[
1+
\frac{
N^{BS}P_{\max}\Gamma_R^{\max}\left(
1-\mathcal O\left((N^R)^{-1}\right)
\right)
}{
\sigma^2B_{\max}
}
\Bigg]
\Bigg), \nonumber
\end{equation}    
\end{small}
where $\Gamma_R^{\max}$ denotes the finite effective
hybrid-channel gain under the ideal continuous passive-reflection environment. Therefore, BSs provide the dominant active capacity-bearing scaling through $N^{BS}$, while RISs contribute a bounded passive channel-gain correction inside the logarithmic capacity expression.

The following proposition gives the exact IREE gain condition when an additional BS or RIS is deployed.

\begin{Prop}[IREE Marginal Gain Condition]
\label{prop:IREE_gain_condition}
Consider a network state with aggregate capacity $C_{Tot}$, mismatch $\xi$, and power consumption $P_T$. After adding one BS or one RIS, denote the corresponding quantities by $C_{Tot}^{+}$, $\xi^{+}$, and $P_T^{+}$. The new deployment improves IREE if and only if
\begin{eqnarray}
\frac{
\min\{C_{Tot}^{+},D_{Tot}\}
}{
\min\{C_{Tot},D_{Tot}\}
}
\left[
1+
\frac{\xi-\xi^{+}}{1-\xi}
\right]
>
\frac{P_T^{+}}{P_T}.
\label{eqn:IREE_gain_condition}
\end{eqnarray}
\end{Prop}

\IEEEproof
This condition follows directly by substituting the pre- and post-deployment IREE values into Definition~\ref{def:IREE} and rearranging the resulting inequality $\eta_{IREE}^{+}>\eta_{IREE}$.
\endIEEEproof

Proposition~\ref{prop:IREE_gain_condition} shows that an additional infrastructure element is useful only when its capacity gain and mismatch-reduction gain jointly compensate for its power penalty.

\subsection{Traffic-Capacity Mismatch Scaling}
\label{subsec:mismatch_scaling_new}

We next characterize how the traffic-capacity mismatch decreases as additional BSs and RISs are deployed. Define the normalized traffic and capacity densities as
\begin{eqnarray}
f_d(\mathcal L)=
\frac{D_T(\mathcal L)}{D_{Tot}},
\quad
f_c(\mathcal L)=
\frac{C_S(\mathcal L)}{C_{S,Tot}},
\label{eqn:normalized_density_scaling}
\end{eqnarray}
where
$C_{S,Tot}=
\iint_{\mathcal A}C_S(\mathcal L)\mathrm d\mathcal L$.
Since $C_S(\mathcal L)$ is the capacity lower bound represented by the ADD-RBF architecture, the following analysis considers the order-wise behavior of $\xi(C_S,D_T)$.

BSs and RISs correct the spatial mismatch through different physical mechanisms. Additional BSs introduce new active transmission sources and progressively establish and reshape the large-scale capacity footprint over the service region. Once this active-capacity footprint has been sufficiently aligned with the main spatial variation of the traffic demand, the remaining mismatch is mainly associated with finer spatial variations within the BS-illuminated regions. In the considered large-element regime, the phase-controlled RIS-assisted atoms can provide a finer spatial resolution than the BS direct-path atoms and are therefore particularly suitable for refining this post-BS residual mismatch. The corresponding BS and RIS resolution conditions are specified in Appendix~\ref{appendix:mismatch_scaling}.

The post-BS residual may remain spatially diffuse over an extended region or become concentrated around a fixed number of localized hotspots. These two residual morphologies lead to different coverage and refinement requirements for
additional RIS deployment.

\begin{Thm}[Residual-Structure-Dependent Mismatch Scaling]
\label{thm:mismatch_scaling_new}
Consider increasing numbers of optimized BSs and RISs. Let $t_g$ characterize the spatial regularity of the large-scale traffic-capacity structure corrected by BS densification, and let $t_\ell$ characterize the regularity of a spatially diffuse post-BS residual. 
Suppose that the residual mismatch is either spatially diffuse over a two-dimensional region of non-vanishing area or hotspot-dominated, with the dominant mismatch concentrated within a fixed collection of localized regions. Under the finite-resolution and continuous RIS-synthesis regularity conditions specified in Appendix~\ref{appendix:mismatch_scaling}, the best-achievable mismatch satisfies
\begin{eqnarray}
\xi(C_S,D_T)=
\mathcal O
\left(
(N^{BS})^{-t_g/2}
+
\mathcal E_\ell^R(N^R)
\right),
\label{eqn:mismatch_scaling_new}
\end{eqnarray}
where the RIS-side local-residual approximation error follows
\begin{small}
\begin{eqnarray} 
\mathcal E_\ell^R(N^R) = 
\begin{cases}
\mathcal O\left(
(N^R)^{-t_\ell/2}
\right),
& \text{spatially diffuse residual}, \\
\mathcal O\left(
\exp[-c_R\sqrt{N^R}]
\right),
&
\text{hotspot-dominated residual},
\end{cases} \nonumber
\end{eqnarray}
\end{small}
where $c_R>0$ is independent of $N^R$ and characterizes the high-order discretization rate of the continuous RIS synthesis over the fixed hotspot regions.
\end{Thm}

\IEEEproof
Please refer to Appendix~\ref{appendix:mismatch_scaling}.
\endIEEEproof

\begin{figure}[t]
\centering
\subfigure[Urban traffic residual]{
\begin{minipage}[c]{0.45\linewidth}
\centering
\includegraphics[height=3.6cm]{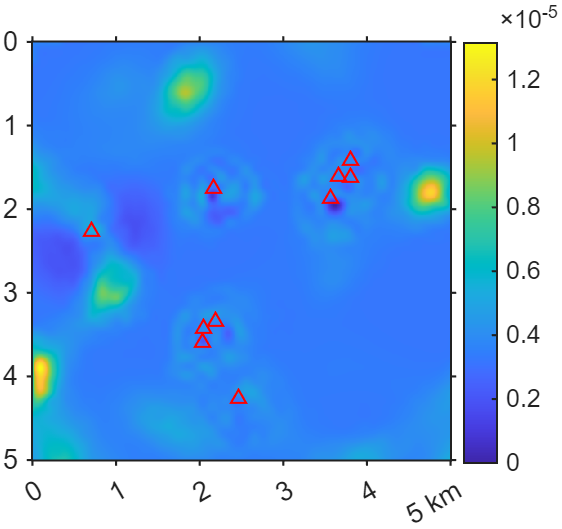}
\label{fig:urban_residual}
\end{minipage}}
\hspace{0.1cm}
\subfigure[Rural traffic residual]{
\begin{minipage}[c]{0.45\linewidth}
\centering
\includegraphics[height=3.6cm]{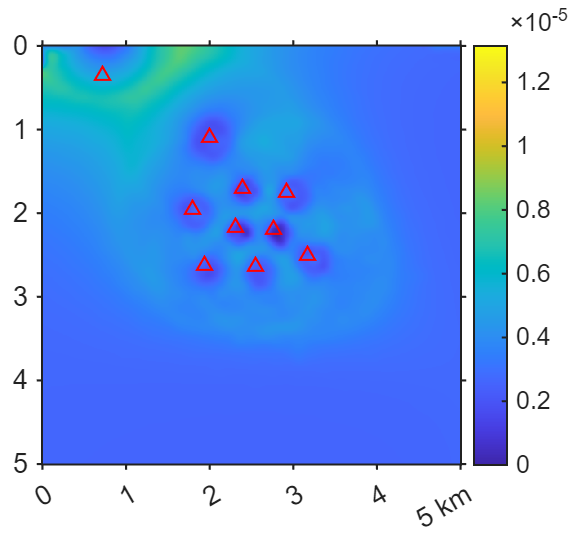}
\label{fig:rural_residual}
\end{minipage}}
\par\medskip
\subfigure[Relative JS divergence versus $N^{R}$]{
\begin{minipage}[c]{0.95\linewidth}
\centering
\includegraphics[width=7.5cm, height=3.6cm]{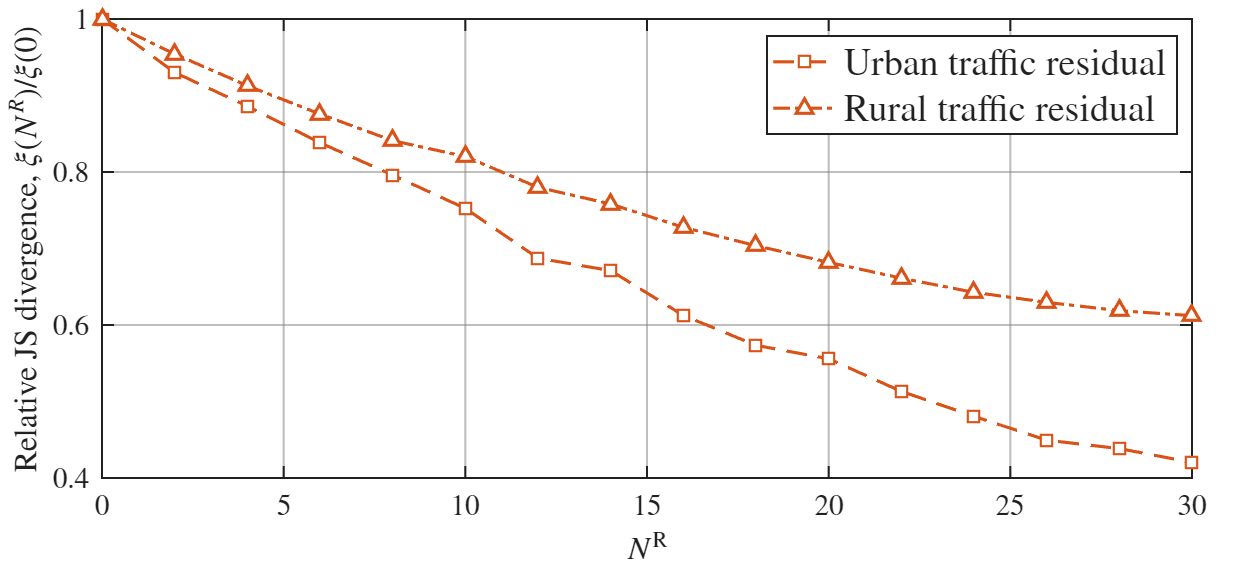}
\label{fig:JS_NR}
\end{minipage}}
\caption{
Illustration of the residual-structure-dependent RIS mismatch reduction in Theorem~\ref{thm:mismatch_scaling_new}, where red
triangles indicate BS positions. In (c), the JS divergence of
each traffic scenario is normalized by its corresponding no-RIS value, i.e., $\xi_s(N^R)/\xi_s(0)$,
$s\in\{\mathrm{urban},\mathrm{rural}\}$. Simulation settings are given in Section~\ref{sect:num_res} with $N^{BS}=10$.
}
\label{fig:residual_scaling}
\end{figure}

Theorem~\ref{thm:mismatch_scaling_new} admits a direct coverage--refinement interpretation. The BS-side term is polynomial because additional BSs must progressively reduce the spacing between active-capacity footprints over a two-dimensional service region. For a spatially diffuse residual, RIS deployment remains coverage-limited: additional RISs must also be distributed over an extended region, and the resulting mismatch reduction is therefore polynomial. For a hotspot-dominated residual, however, the relevant spatial supports remain fixed, and additional RISs are mainly used to refine the same localized capacity profiles. This repeated refinement yields the stretched-exponential term $\mathcal O(\exp[-c_R\sqrt{N^R}])$. To compare the mismatch-reduction rates under different initial mismatch levels, Fig.~\ref{fig:residual_scaling} plots the JS divergence of each scenario relative to its corresponding no-RIS value. Over the considered deployment range, the hotspot-dominated urban residual exhibits a faster relative reduction than the spatially diffuse rural residual, illustrating the practical distinction between localized refinement and extended-area coverage.


\subsection{IREE-Oriented Deployment Regimes}
\label{subsec:IREE_deployment_regimes}

We finally translate the preceding scaling results into IREE-oriented deployment guidelines. Proposition~\ref{prop:IREE_gain_condition} provides the exact finite-deployment condition for determining whether an additional BS or RIS improves IREE, while Theorem~\ref{thm:mismatch_scaling_new} identifies the spatial mechanisms governing the corresponding mismatch-reduction gain. Based on the dominant source of IREE loss, we distinguish the capacity-limited and mismatch-limited regimes.

In the capacity-limited regime, the aggregate capacity remains below the traffic demand. For a marginal deployment that keeps both the current and updated networks capacity-limited, i.e.,
$C_{Tot}\leq D_{Tot}$ and $C_{Tot}^{+}\leq D_{Tot}$,
\eqref{eqn:IREE_gain_condition} reduces to
\begin{eqnarray}
\frac{C_{Tot}^{+}}{C_{Tot}}
\left[
1+
\frac{\xi-\xi^{+}}{1-\xi}
\right]
>
\frac{P_T^{+}}{P_T}.
\label{eqn:capacity_limited_gain_condition}
\end{eqnarray}
Here, an additional infrastructure element must provide sufficient aggregate-capacity and mismatch-reduction gains to compensate for its power penalty. BS deployment is generally favored in this regime because BS densification provides the dominant active capacity-bearing scaling in $C_{Tot}$ while simultaneously reshaping the large-scale capacity footprint over the service region. By contrast, RIS deployment provides only a bounded passive channel-gain correction and cannot offer the same active-capacity scaling as additional BSs. RISs may still improve IREE when the initial capacity deficit is small and their passive gain is sufficiently large, but the preferred deployment must ultimately satisfy \eqref{eqn:capacity_limited_gain_condition}. If the marginal deployment crosses the boundary $C_{Tot}^{+}>D_{Tot}$, the general condition in \eqref{eqn:IREE_gain_condition} should be used directly.

In the mismatch-limited regime, the aggregate traffic demand has already been supported, i.e., $C_{Tot}>D_{Tot}$, while the remaining IREE loss is mainly caused by spatial traffic-capacity mismatch. For a marginal deployment that keeps the network in this regime, the aggregate-capacity factor in the IREE numerator remains saturated at $D_{Tot}$, and \eqref{eqn:IREE_gain_condition} becomes
\begin{eqnarray}
\xi-\xi^{+}
>
(1-\xi)
\frac{P_T^{+}-P_T}{P_T}.
\label{eqn:mismatch_limited_gain_condition}
\end{eqnarray}
Therefore, additional infrastructure is beneficial only when its mismatch-reduction gain exceeds the corresponding relative power penalty. If the large-scale active-capacity footprint remains poorly aligned with the traffic distribution, additional BS deployment can still be effective because it can reshape this footprint while maintaining active transmission capability. Once the large-scale footprint has been sufficiently established and the post-BS residual mismatch becomes dominant, RIS deployment becomes increasingly attractive because its phase-controlled reflected paths can provide finer spatial refinement without introducing another active transmission source. The direct hardware-power increment of an RIS is only its circuit power $P^r$, whereas an additional BS introduces both circuit power and active transmission capability. The actual total-power change after network reoptimization is nevertheless fully captured by $P_T^{+}-P_T$ in \eqref{eqn:mismatch_limited_gain_condition}.

The spatial morphology of the post-BS residual further affects the mismatch-reduction gain in~\eqref{eqn:mismatch_limited_gain_condition}. For a spatially diffuse residual, both BS and RIS densification exhibit polynomially diminishing gains, so their asymptotic orders alone do not determine an unconditional winner; the preferred deployment still depends on the mismatch reduction and the corresponding power increment. Nevertheless, once sufficient BS illumination has been established, the lower hardware-power cost of an RIS can make it more likely to satisfy the gain condition. For a hotspot-dominated residual, additional RISs can repeatedly refine a fixed collection of localized capacity profiles, leading to the faster mismatch reduction characterized in Theorem~\ref{thm:mismatch_scaling_new}. RIS deployment is therefore more favorable in hotspot-dominated mismatch-limited networks, while the final decision remains subject to the joint mismatch and power tradeoff.

The resulting deployment guideline is thus regime dependent. BS deployment should generally be prioritized when insufficient aggregate capacity or a poorly established large-scale capacity footprint is the dominant bottleneck. RIS deployment becomes increasingly favorable when the network is mismatch-limited and the remaining spatial mismatch can be efficiently refined within sufficiently illuminated regions, especially when it is concentrated around localized traffic hotspots.

\section{Numerical Results} \label{sect:num_res}

\begin{figure}[t]
\centering
\subfigure[Urban traffic profile with 
$(\mu_{loc},\sigma_{scl},w_{\max},d_{coh})$ given by $(19,2.4,0.003,71.4{\rm m})$.]{
\begin{minipage}[c]{0.45\linewidth}
\centering
\includegraphics[height=3.6cm]{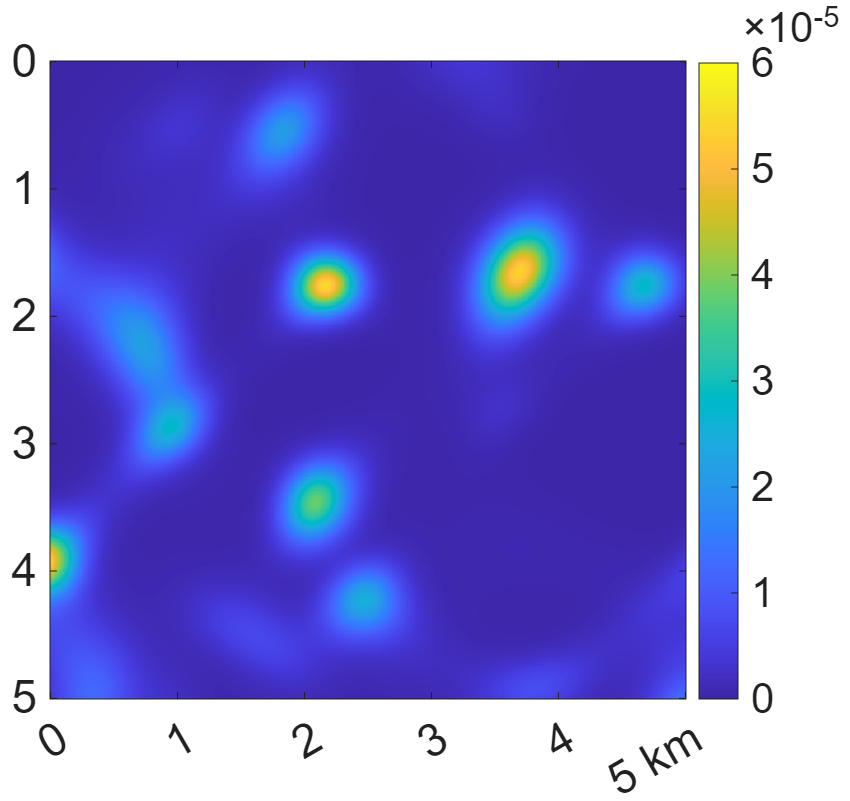}
\label{fig:traffic_urban}
\end{minipage}}
\hspace{0.1cm}
\subfigure[Rural traffic profile with 
$(\mu_{loc},\sigma_{scl},w_{\max},d_{coh})$ given by $(12.5,2.8,0.0012,1075.4{\rm m})$.]{
\begin{minipage}[c]{0.45\linewidth}
\centering
\includegraphics[height=3.6cm]{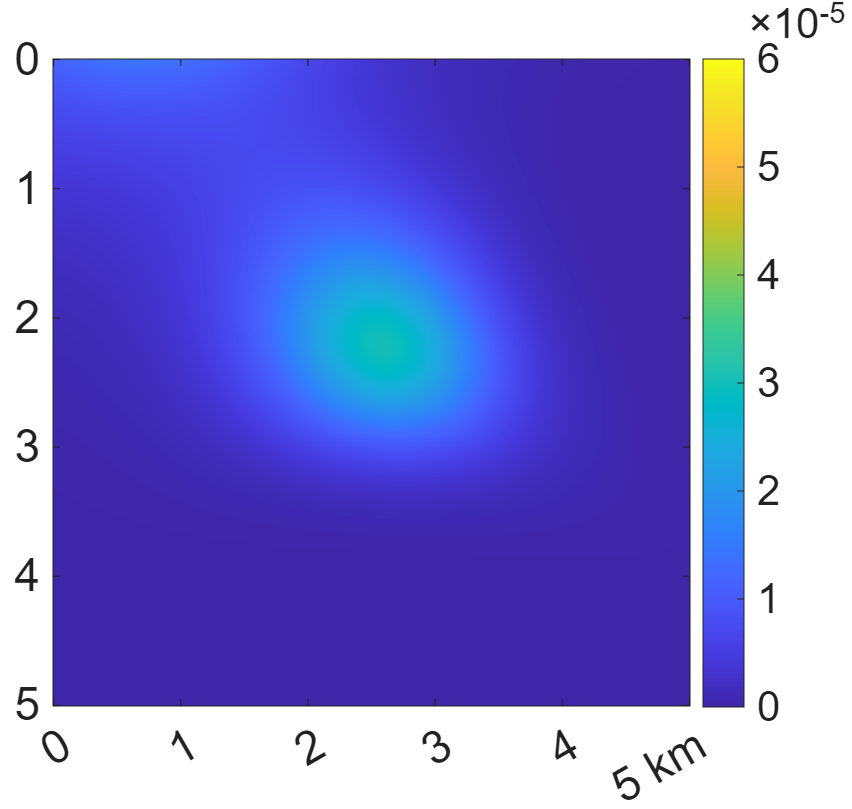}
\label{fig:traffic_rural}
\end{minipage}}
\caption{Representative urban and rural traffic profiles generated using the measurement-driven spatial traffic-density model in \cite{lee2014spatial}.}
\label{fig:traffic}
\end{figure}

\begin{table} [t] 
\centering 
\caption{Simulation Parameters}  
\label{tab:simu_para}
\footnotesize
\begin{tabular}{c | c }  

\toprule
Number of BSs, $N^{BS}$ & $36$  \\

\midrule
Number of RISs, $N^{R}$ & $36$  \\

\midrule
Number of antennas, $N^{BS}_T$ & $16$  \\

\midrule
Number of elements, $N^R_T$ & $100$  \\


\midrule
Height of BSs & $25$ m \\

\midrule
Maximum Power, $P_{\max}$ & $60$ W \\

\midrule
Carrier frequency, $F_c$ & $6.775$ GHz \\

\midrule
Aggregate bandwidth budget, $B_{\max}$ & $6$ GHz \\

\midrule
Minimum CSAT, $\zeta_{\min}$ & $0.8$ \\

\midrule  
Circuit power of BS, $P^{c}$ & $100$ W \cite{xu2012energy} \\

\midrule  
Circuit power of RIS, $P^{r}$ & $3$ W \\

\midrule  
Efficiency of power amplifier, $1/ \lambda $ & $38\%$ \\

\midrule 
Standard deviation of shadowing & $10$ dB \\

\midrule 
Path loss (dB) & $28 + 22\log_{10}d + 20\log_{10}(F_c)$ \\

\midrule 
Power spectral density of noise, $\sigma^2$ & $-174$ dBm/Hz \\

\midrule 
Aggregate traffic demand, $D_{Tot}$ & $100$ Gbit/s \\

\midrule 
Traffic distribution profile & Urban traffic in Fig.~\ref{fig:traffic_urban} \\

\bottomrule
\end{tabular}  
\end{table}

In this section, we present comprehensive numerical simulations to evaluate the performance of the proposed ADD-RBF scheme and validate the derived IREE scaling. The simulations consider both urban and rural traffic profiles to assess the effectiveness of joint multi-BS and multi-RIS deployment under spatially heterogeneous conditions. The results corroborate the theoretical scaling principles and provide practical guidelines for green 6G network design.

\begin{figure*}[t] 
\centering  
\includegraphics[height=8cm,width=18cm]{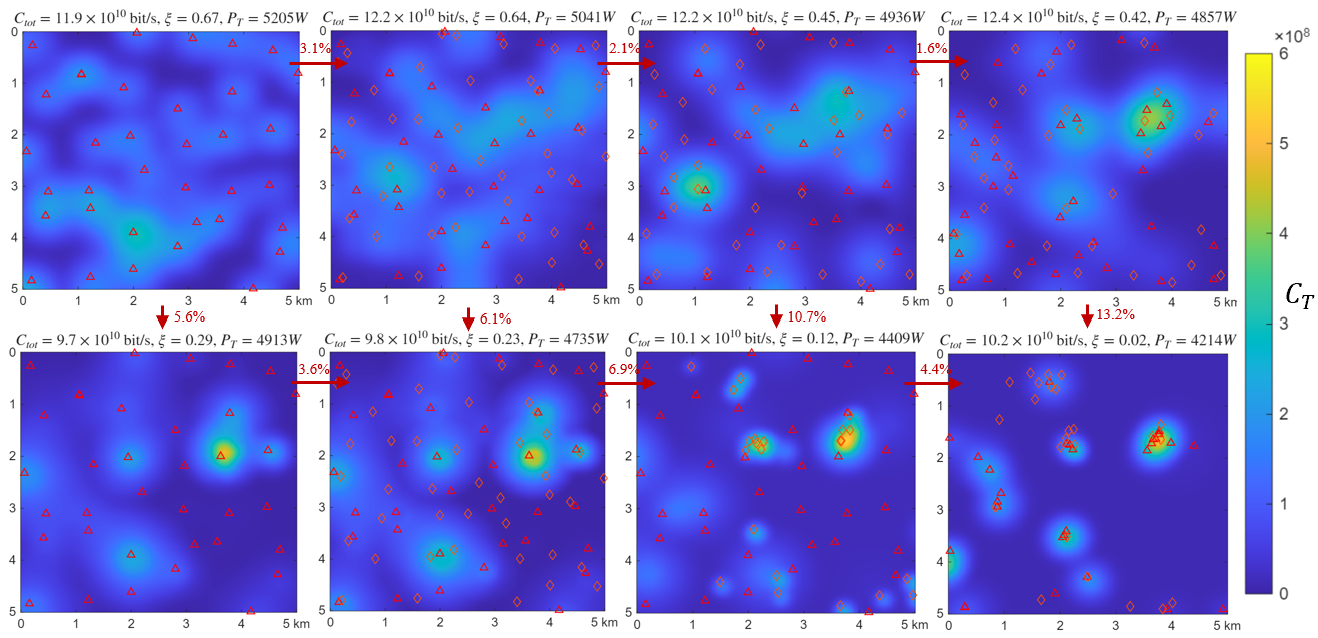}
\caption{Comparative evaluation across four network deployment configurations. The top row displays the performance achieved by the baseline scheme, while the bottom row corresponds to results from the proposed ADD-RBF scheme. From left to right, the first column depicts deployments using BSs only with unoptimized locations; the second column shows joint BS and RIS deployment where the positions are not optimized; the third column represents joint deployments featuring optimized RIS placements alongside unoptimized BS locations; and the fourth column showcases the scenario with both BS and RIS locations optimized. Red triangles indicate BS positions, and orange diamonds denote RIS placements. }
\label{fig:baseline_simu}
\end{figure*}

In the following simulations, we consider a $5 \times 5$ km square area. The adopted path-loss model uses a representative 3GPP TR~38.901 UMa LOS path-loss expression \cite{3gpp_tr38901}, with $d$ in meters and $F_c$ in GHz. The traffic demand is modeled by two separate components: the aggregate traffic demand $D_{Tot}$ over the whole area $\mathcal A$ and the normalized spatial traffic-density profile. The former specifies the total traffic demand in the considered service area, while the latter determines how this traffic demand is spatially distributed. To evaluate the proposed deployment framework under different spatial mismatch patterns, we consider two representative profiles, namely an urban traffic profile and a rural traffic profile, as illustrated in Fig.~\ref{fig:traffic}. The detailed simulation parameters, unless otherwise specified, are provided in Table~\ref{tab:simu_para}.

The two traffic profiles in Fig.~\ref{fig:traffic} are generated using the measurement-driven spatial traffic-density model in \cite{lee2014spatial}, which represents cellular traffic as a log-normal transformation of a spatially correlated Gaussian random field. The parameters $\mu_{loc}$ and $\sigma_{scl}$ determine the marginal statistics of the unnormalized traffic density, while $w_{\max}$ and $d_{coh}$ characterize its spatial fluctuation and correlation range, respectively. After generation, each traffic map is normalized and scaled to the prescribed aggregate demand $D_{Tot}$, so these parameters mainly determine the spatial traffic pattern rather than the total load. Accordingly, the urban profile in Fig.~\ref{fig:traffic_urban}, with a larger $\mu_{loc}$ and much shorter $d_{coh}$, exhibits stronger localized hotspots and faster spatial variations, whereas the rural profile in Fig.~\ref{fig:traffic_rural}, with a smaller $\mu_{loc}$, smaller $w_{\max}$, and much longer $d_{coh}$, is smoother and more spatially continuous. These two profiles are used to evaluate the proposed framework under fine-grained urban mismatch and smooth rural mismatch.

\subsection{IREE Comparison with Baselines}

In Fig.~\ref{fig:baseline_simu}, we conduct a comparative study for a joint multi-BS and multi-RIS deployment scenario. For the baseline, the locations of both BSs and RISs are determined through the K-means algorithm \cite{chen2022allocation}, while beamforming vectors at the BSs are constructed based on the zero-forcing technique \cite{spencer2004zero}, and RIS phase shifts are optimized via the phase alignment method \cite{8647620}. 

The performance achieved by introducing RISs is examined under progressively refined deployment strategies. As shown in Fig.~\ref{fig:baseline_simu}, introducing RISs with unoptimized locations, subsequently optimizing the RIS locations, and further co-optimizing both BS and RIS locations lead to consistent reductions in JS divergence. This mitigation of spatial mismatch results in power consumption reductions of $3.6\%$, $6.9\%$, and $4.4\%$, respectively, translating into IREE gains of $15.2\%$, $23.6\%$, and $16.5\%$. These results underscore the exceptional capability of RISs in capturing spatial traffic correlations and alleviating localized hotspots, as highlighted in the theoretical analysis. By adaptively aligning capacity with heterogeneous demand, RIS integration significantly enhances the energy-saving potential of the network.

Furthermore, to evaluate the effectiveness of the proposed ADD-RBF scheme, we compare its performance against the baseline approach. The observed superiority of ADD-RBF is consistent with the representational flexibility of the dual-RBF architecture established in Theorem~\ref{thm:Existence_risRBF}.  This theoretical guarantee enables the framework to directly reduce JS divergence by aligning capacity provision with spatial traffic demand. Experimentally, ADD-RBF achieves further JS divergence reduction alongside power savings of $5.6\%$, $6.1\%$, $10.7\%$, and $13.2\%$, culminating in remarkable IREE gains of $121.1\%$, $126.1\%$, $79.1\%$, and $94.7\%$, respectively. These results confirm that the proposed scheme effectively shifts the optimization focus toward holistic traffic-capacity mismatch mitigation while maintaining high energy efficiency, leading to more sustainable and green network operation.

\begin{figure*} [t]
\centering
\subfigure[IREE versus number of BSs and RISs.]{
\begin{minipage}[c]{0.45\linewidth}
\centering
\includegraphics[height=7cm,width=8cm]{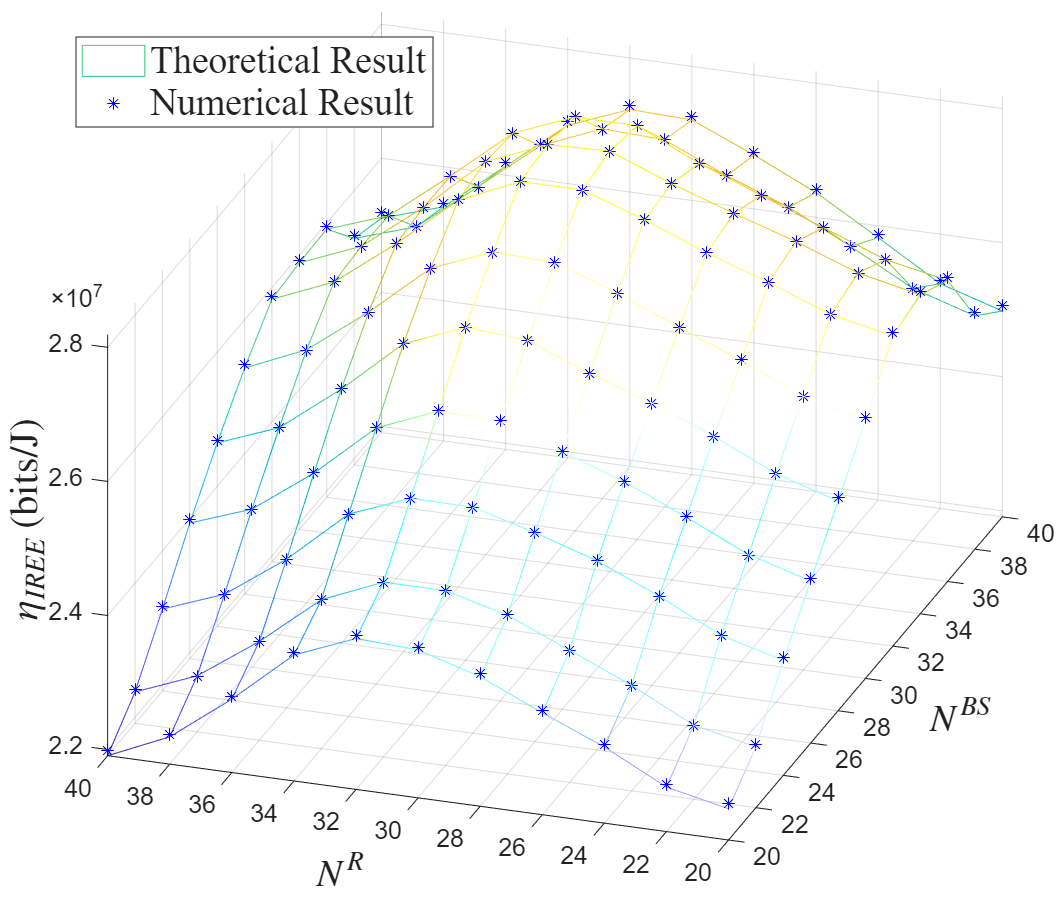}
\label{fig:IREE3D}
\end{minipage}}
\subfigure[Gradient field of IREE based on numerical results.]{
\begin{minipage}[c]{0.45\linewidth}
\centering
\includegraphics[height=7cm,width=7.5cm]{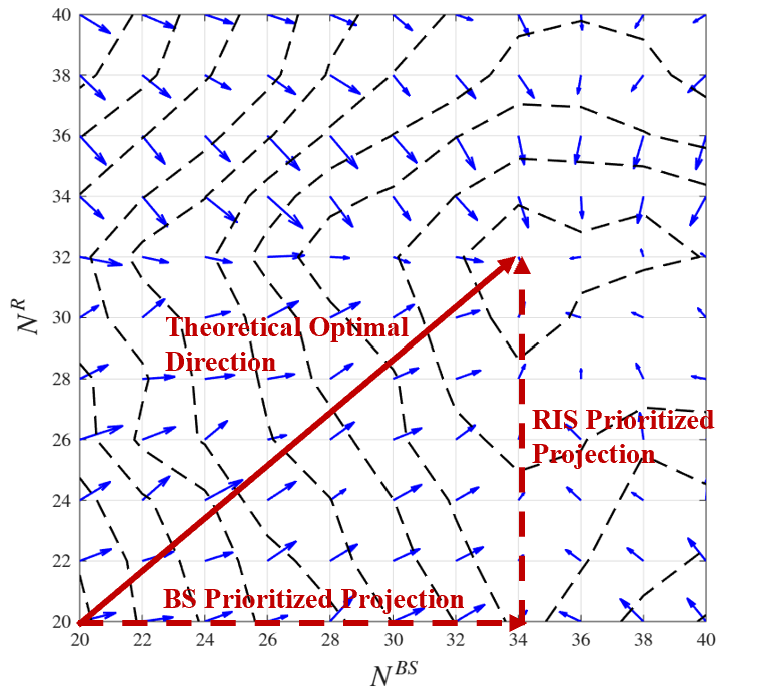}
\label{fig:IREE_gradient}
\end{minipage}}
\subfigure[Aggregate capacity \& JS divergence versus changes in the numbers of BSs or RISs with initial settings $(N^{BS},N^{R}) = (20, 20)$. ]{
\begin{minipage}[c]{0.45\linewidth}
\centering
\includegraphics[height=8cm,width=8cm]{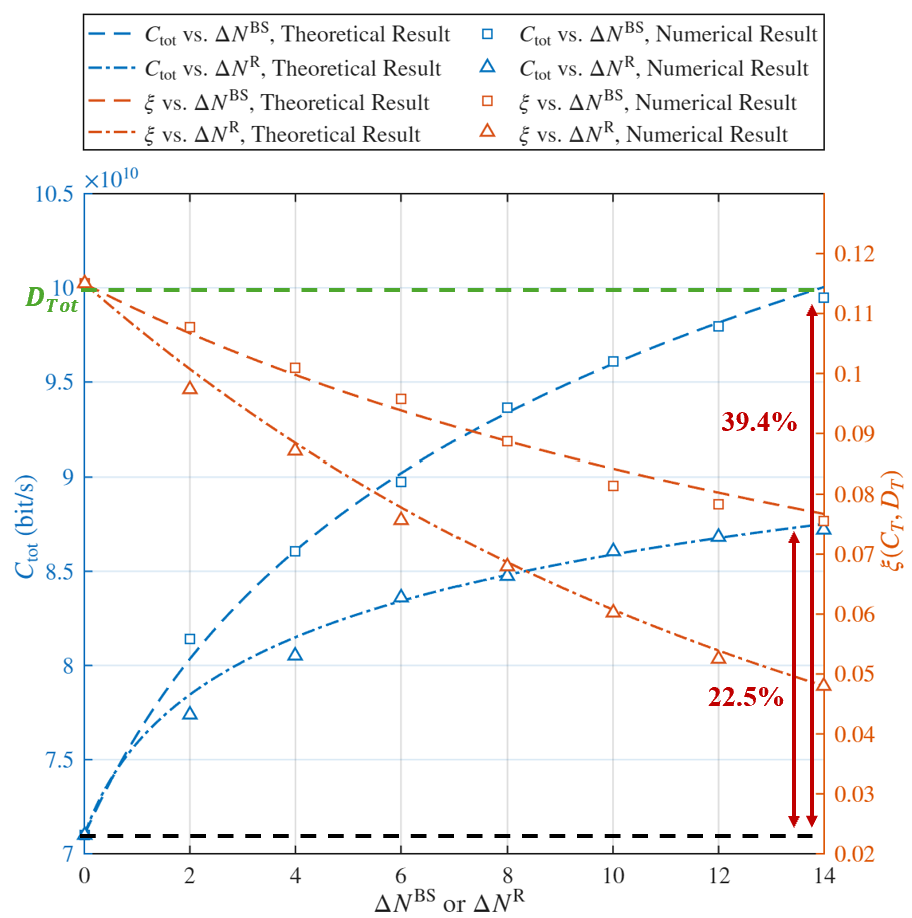}
\label{fig:capacity_scale_row}
\end{minipage}}
\subfigure[Aggregate capacity \& JS divergence versus changes in the numbers of BSs or RISs with initial settings $(N^{BS},N^{R}) = (34, 20)$. ]{
\begin{minipage}[c]{0.45\linewidth}
\centering
\includegraphics[height=8cm,width=8cm]{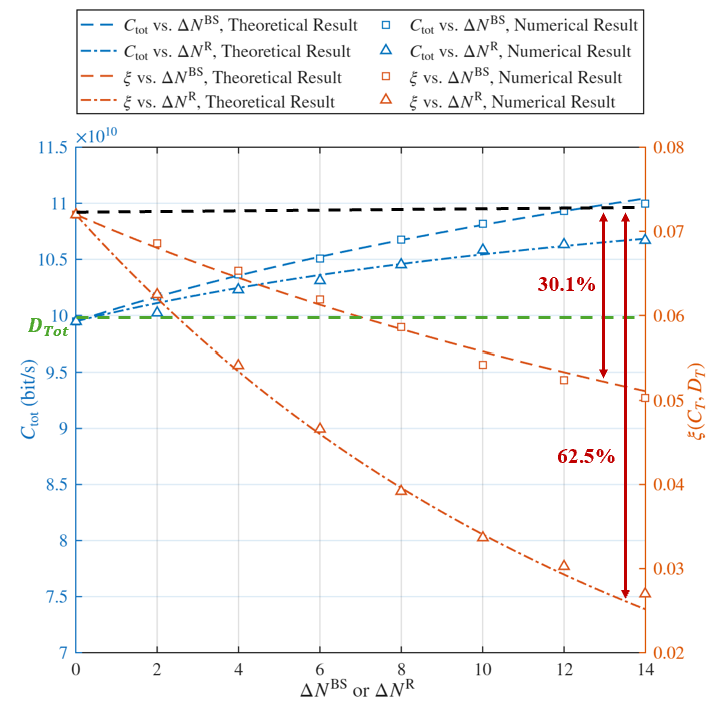}
\label{fig:capacity_scale_column}
\end{minipage}}
\caption{Numerical illustration of IREE behavior under joint BS/RIS deployment.}
\label{fig:simu_scaling}
\end{figure*}

\subsection{Performance Scaling with Different Numbers of BSs \& RISs}

Fig.~\ref{fig:simu_scaling} shows the IREE scaling behavior as the numbers of BSs and RISs increase. The theoretical results are constructed component-wise from the scaling laws derived in Section~\ref{sect:scaling_law}. Specifically, the aggregate-capacity realization preserves the logarithmic BS growth and the first-order saturation gap
of the RIS-assisted gain. Since the considered urban traffic produces a hotspot-dominated post-BS residual, the mismatch realization adopts the corresponding branch of Theorem~\ref{thm:mismatch_scaling_new}, $\widehat{\xi} = A_B \left( N^{BS} + \upsilon_B \right)^{-t_g/2} + A_R \exp \left[ -c_R\sqrt{N^R+\upsilon_R} \right]$. The regularity order $t_g=1.07$ is independently obtained from a log--log regression of the third-order spatial increments of the coarse-scale traffic component and is fixed before the finite-range calibration. The remaining coefficients of the capacity and mismatch realizations are jointly calibrated over the considered deployment range, with one common coefficient set shared by the BS- and RIS-scaling directions of each component. The resulting capacity and
mismatch realizations are then combined with the power-consumption model in \eqref{eqn:total_pow_def} and the
IREE definition in \eqref{eq:def_iree} to reconstruct the theory-guided IREE surface.

As shown in Fig.~\ref{fig:IREE3D}, the reconstructed theoretical surface closely follows the numerical IREE results over the considered deployment range, showing that the fitted scaling forms provide a useful description of the observed numerical trends. The gradient field in Fig.~\ref{fig:IREE_gradient} further shows that the relative marginal benefits of BS and RIS deployment vary across the deployment region. Consequently, the preferred expansion direction depends on the current operating point, consistent with the marginal IREE-gain condition in Proposition~\ref{prop:IREE_gain_condition}.

Fig.~\ref{fig:capacity_scale_row} considers the initial deployment $(N^{BS},N^R)=(20,20)$, where $C_{Tot}\leq D_{Tot}$ and the network is capacity-limited. Increasing the number of BSs by 14 raises the aggregate capacity by $39.4\%$ and improves IREE by $27.1\%$. By comparison, adding 14 RISs increases the aggregate capacity by $22.5\%$ and improves IREE by $11.4\%$. Although RIS deployment provides a faster reduction in JS divergence, the IREE numerator remains limited by the available aggregate capacity. BS densification is therefore more
effective at this operating point because it provides a stronger active-capacity gain while also reducing the spatial mismatch.

The dominant mechanism changes in Fig.~\ref{fig:capacity_scale_column}, where the initial deployment is $(N^{BS},N^R)=(34,20)$ and $C_{Tot}>D_{Tot}$. Adding 14 RISs reduces the JS divergence by $62.5\%$ and improves IREE by $12.5\%$, whereas adding the same number of BSs reduces the JS divergence by only $30.1\%$ and decreases IREE by $17.4\%$. Once sufficient aggregate capacity has been
established, further BS densification provides diminishing
mismatch-reduction benefits while incurring a relatively large power increment. In contrast, RISs can refine the remaining local traffic-capacity mismatch with a substantially smaller circuit-power cost. These results are consistent with the regime-dependent deployment discussion in Section~\ref{subsec:IREE_deployment_regimes}: BS deployment tends to be more effective when aggregate capacity is insufficient, whereas RIS deployment can become more attractive once sufficient active capacity has been established and spatial mismatch becomes a major source of the remaining IREE loss.

\begin{figure*} [t]
\centering
\subfigure[IREE versus number of BSs under $N^{R} = 32$.]{
\begin{minipage}[c]{0.45\linewidth}
\centering
\includegraphics[height=8cm,width=7.5cm]{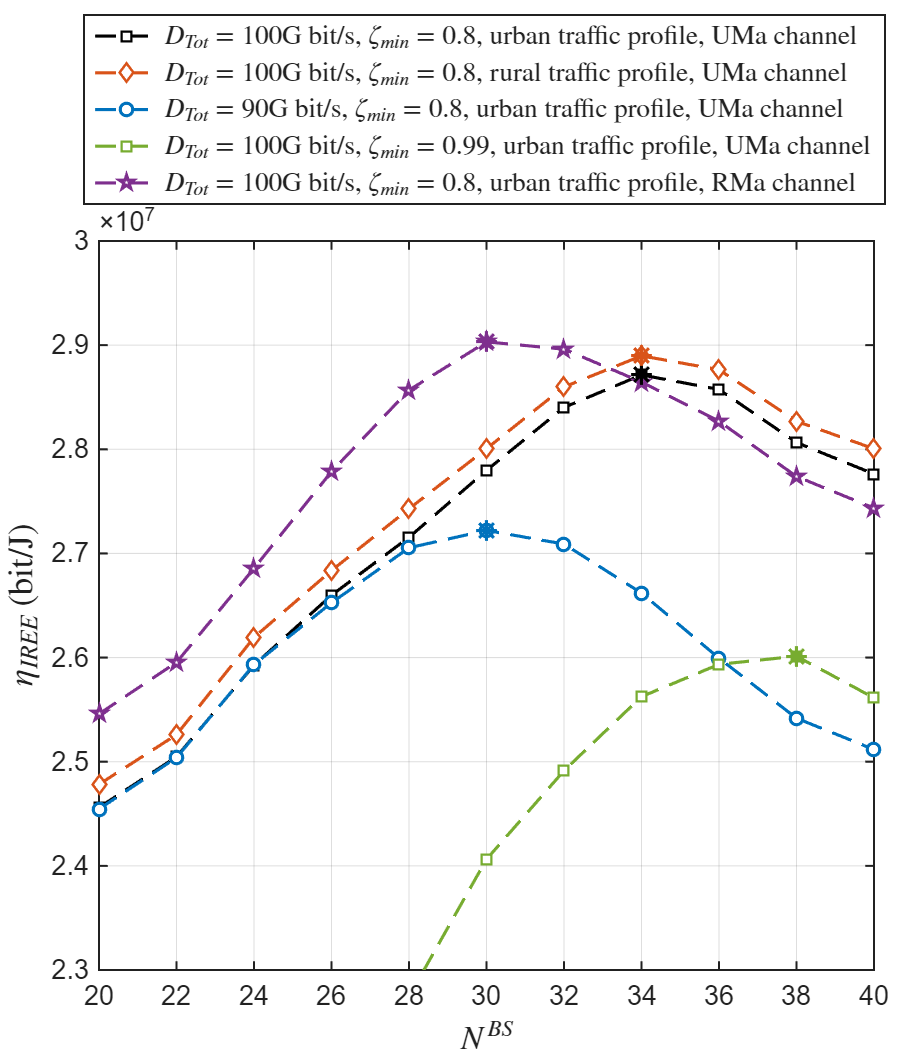}
\label{fig:IREE_NBS}
\end{minipage}}
\subfigure[IREE versus number of RISs under $N^{BS} = 34$.]{
\begin{minipage}[c]{0.45\linewidth}
\centering
\includegraphics[height=8cm,width=7.5cm]{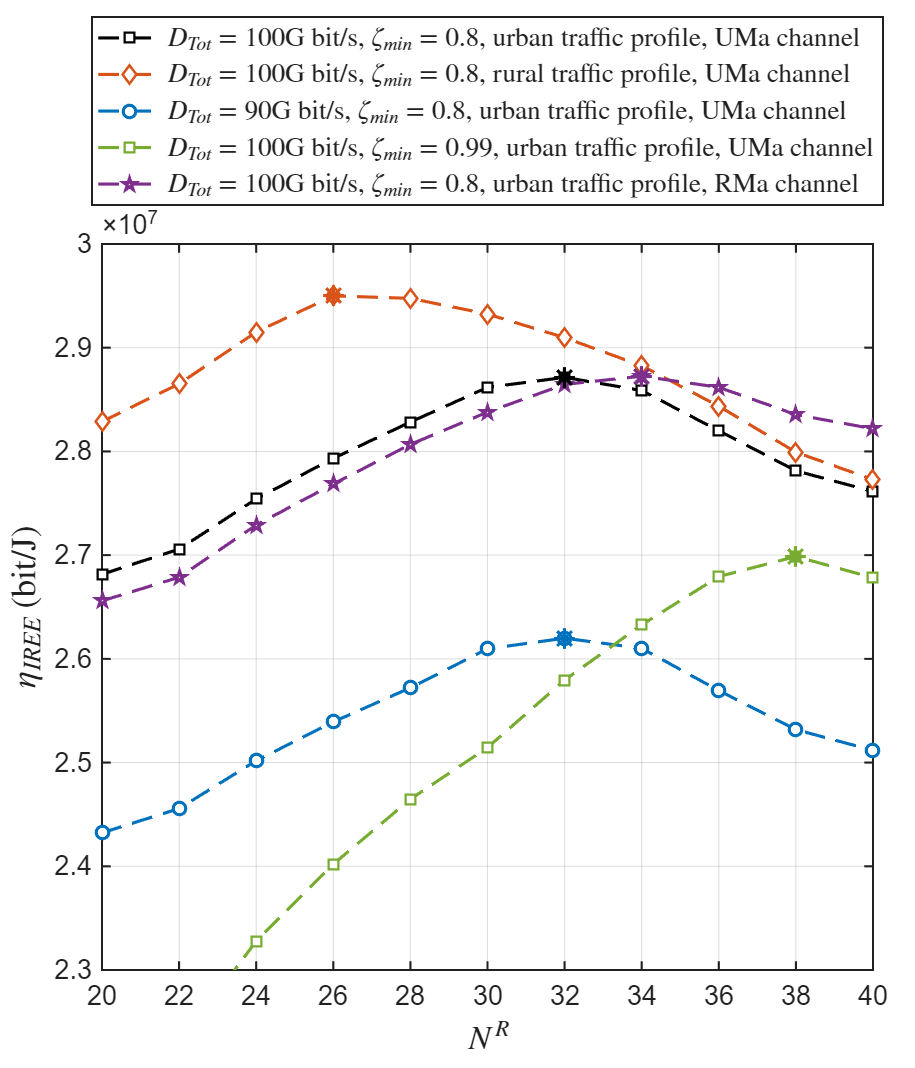}
\label{fig:IREE_NR}
\end{minipage}}
\caption{IREE sensitivity to the spatial traffic profile, aggregate traffic demand, propagation model, and CSAT requirement.}
\label{fig:design_principle}
\end{figure*}

\subsection{Design Principle}

Fig.~\ref{fig:design_principle} reveals a clear difference between the factors governing BS and RIS deployment. Under the same aggregate traffic demand, replacing the urban traffic profile with the smoother rural profile has little effect on the IREE-maximizing number of BSs in Fig.~\ref{fig:IREE_NBS}, but shifts the optimal number of RISs in Fig.~\ref{fig:IREE_NR} from approximately $32$ to $26$. The active-capacity requirement is mainly determined by the aggregate traffic load, whereas the smoother rural profile requires less RIS-assisted spatial refinement than the more spatially heterogeneous urban profile.

This distinction is further supported by the reduced-traffic case. When $D_{Tot}$ is decreased while the normalized traffic profile remains unchanged, the optimal number of BSs decreases from approximately $34$ to $30$, whereas the optimal RIS number changes only slightly. Therefore, the aggregate traffic volume primarily determines the required BS scale, while the spatial heterogeneity of the traffic demand primarily determines the RIS refinement scale.

The channel model can further alter the deployment balance. To illustrate this effect, we replace the default UMa channel with the RMa channel, given by ${\rm PL}_{\rm RMa}^{\rm LOS} = 0.001398d + 20\log_{10}(40\pi d F_c/3) + 0.5\log_{10}d - 0.7$ \cite{3gpp_tr38901}, while retaining the urban traffic profile and all other system settings. Over the considered service range, the RMa channel exhibits weaker distance-dependent attenuation than the UMa channel, enabling each BS to provide higher capacity under the same transmit-power budget and thereby reducing the required BS scale. Meanwhile, the resulting BS-induced capacity distribution varies more slowly in space, making slightly more RIS-assisted refinement beneficial for matching the localized traffic fluctuations. As shown in Fig.~\ref{fig:design_principle}, adopting the RMa channel reduces the optimal number of BSs from $34$ to $30$, while increasing the optimal number of RISs from $32$ to $34$.

Finally, increasing the CSAT requirement $\zeta_{\min}$ shifts the preferred
deployment toward larger numbers of both BSs and RISs, since a stricter $\zeta_{\min}$ calls for both sufficient aggregate capacity and closer spatial traffic--capacity alignment. Overall, the results suggest a simple deployment guideline: traffic volume governs how much active BS capacity is needed, whereas traffic heterogeneity governs how much passive RIS refinement is beneficial.

\section{Conclusion} \label{sect:conc}

In this paper, we developed an ADD-RBF framework that represents the BS direct-path and RIS-assisted spatial capacity contributions through a dual RBF architecture and enables accepted alternating optimization for IREE-oriented joint deployment under traffic-capacity mismatch. We further established that BS densification provides logarithmic active-capacity growth $\mathcal{O}(\log N^{BS})$ and polynomial large-scale mismatch reduction $\mathcal{O}\big((N^{BS})^{-t_g/2}\big)$, whereas RIS deployment provides a bounded passive capacity-gain correction and residual-structure-dependent mismatch reduction, which is polynomial for spatially diffuse residuals and follows the stretched-exponential order $\mathcal{O}\big(\exp[-c_R\sqrt{N^R}]\big)$ for hotspot-dominated residuals under the stated regularity conditions. Numerical results illustrate the corresponding deployment tradeoff: BS expansion is generally preferred in capacity-limited networks, whereas RIS expansion becomes more attractive in mismatch-limited networks, with aggregate traffic demand and traffic heterogeneity mainly governing the preferred BS and RIS scales, respectively.

\section*{List of Abbreviations}

\begin{center}
\small
\begin{tabular}{ll}
\toprule
Abbreviation & Definition \\
\midrule
6G & Sixth-generation \\
ADD-RBF & Alternating directional dual radial basis function \\
AWGN & Additive white Gaussian noise \\
BS & Base station \\
CSAT & Customer satisfaction score \\
EE & Energy efficiency \\
IREE & Integrated relative energy efficiency \\
JS & Jensen--Shannon \\
LoS & Line-of-sight \\
MISO & Multiple-input single-output \\
PCP & Poisson cluster process \\
PPP & Poisson point process \\
QoS & Quality of service \\
RBF & Radial basis function \\
RIS & Reconfigurable intelligent surface \\
SNR & Signal-to-noise ratio \\
RMa & Rural macrocell \\
UMa & Urban macrocell \\
URA & Uniform rectangular array \\
\bottomrule
\end{tabular}
\end{center}

\begin{appendices}

\section{Proof of Lemma~\ref{lem:dual-rbf}}
\label{appendix:dual-rbf}

For the BS direct-path component, the path-loss term depends on the user location through $\|\mathcal L^U-\mathcal L_n^B\|$. Hence, with the remaining parameters fixed, the corresponding spatial atom is a radial function centered at $\mathcal L_n^B$, forming the BS direct-path RBF branch \cite{yu2024iree}.

For an effective BS--RIS pair $(n,m)$, the dominant spatial form of the reflected-power contribution can be written as
\begin{eqnarray}
\psi_{n,m}^{R}(\mathcal L^U;\mathbf\Theta_m)
\propto
\frac{
|A_{n,m}(\mathcal L^U;\mathbf\Theta_m)|^2
}{
L(\mathcal L_m^R,\mathcal L_n^B)^2
L(\mathcal L^U,\mathcal L_m^R)^2
},
\label{eqn:RIS_atom_dual_rbf_appendix}
\end{eqnarray}
where the RIS--user path-loss term provides a radial envelope centered at $\mathcal L_m^R$, while $A_{n,m}(\mathcal L^U;\mathbf\Theta_m)$ provides the phase-controlled spatial modulation. Therefore, the RIS-assisted contributions form a reflected RBF-type dictionary.

Consequently, $C_S(\mathcal L^U)$ is represented by a BS direct-path RBF dictionary and an RIS-assisted reflected RBF-type dictionary. Since their parameters belong to the separate BS-side and RIS-side variable blocks, the two dictionaries can be updated alternately, yielding the dual-RBF architecture.

\section{Proof of Theorem~\ref{thm:conver_analy}}
\label{appendix:conver_analy}

For fixed $\kappa$ and fixed $\eta_{IREE}^{(k)}$, the accepted inner alternating updates are evaluated with respect to a fixed loss $L_{err}(\cdot;\kappa)$. Let
$\mathcal X^t=(\mathcal X_B^t,\mathcal X_R^t)$ denote the $t$-th accepted inner full iterate, where one full iterate consists of one BS-side update followed by one RIS-side update. Applying the sufficient-decrease rule successively to the two block updates gives $L_{err}(\mathcal X^{t+1};\kappa)
\leq
L_{err}(\mathcal X^{t};\kappa)
-
\tau\|\mathcal X^{t+1}-\mathcal X^t\|^2$. Since $L_{err}(\cdot;\kappa)$ is lower bounded, the accepted loss sequence is monotone and converges. Summing the above inequality over $t$ gives $\sum_{t=0}^{\infty}
\|\mathcal X^{t+1}-\mathcal X^t\|^2<\infty$, and hence $\|\mathcal X^{t+1}-\mathcal X^t\|\to 0$.

Because the accepted sequence is bounded, it has accumulation points. If an accumulation point were not block-stationary, then there would exist an admissible descent direction for at least one block. Under the first-order consistency of the accepted Adam-generated candidates and the backtracking acceptance rule, a sufficiently small candidate along this direction would satisfy the sufficient-decrease condition, contradicting the vanishing accepted-step property. Therefore, every accumulation point is block-stationary. The argument applies separately to each fixed-$\kappa$ and fixed-$\eta_{IREE}^{(k)}$ training round.

\section{Proof of Capacity Scaling}
\label{appendix:capacity_scaling}

We first define the effective hybrid-channel gain enabled by $N^R$ discrete RIS panels as $\Gamma_R(N^R)
\triangleq \sup_{n,\mathcal L\in\mathcal A} \left\|
\mathbf h_n(\mathcal L)
\right\|_2^2$. To characterize its dependence on $N^R$, consider an ideal continuous passive-reflection environment over the admissible RIS deployment region.
Since this environment remains passive and the deployment
region, physical apertures, propagation geometry, and link
separation distances are finite, its effective hybrid-channel gain is bounded by a finite constant $\Gamma_R^{\max}$. This quantity represents the maximum utilization of the existing BS transmission capability rather than an additional active capacity source.

A deployment with $N^R$ RIS panels provides a spatial
discretization of this continuous passive environment. Since the admissible two-dimensional deployment region permits a quasi-uniform discretization, its fill distance satisfies $h_R(N^R)=\mathcal O((N^R)^{-1/2})$
\cite{wendland2004scattered,narcowich2006sobolev}. If the
effective channel-gain functional is twice continuously
differentiable around the optimal continuous configuration and its first-order variation vanishes at this optimum, the discretization-induced gain loss is second order in the fill distance. Hence, $0 \leq
\Gamma_R^{\max}-\Gamma_R(N^R)
=\mathcal O\left(h_R(N^R)^2\right) =
\mathcal O\left((N^R)^{-1}\right)$,
or equivalently, $\Gamma_R(N^R)=
\Gamma_R^{\max}
\left[
1-\mathcal O\left((N^R)^{-1}\right)
\right]$. Thus, increasing $N^R$ progressively reduces the spatial discretization gap between the finite RIS deployment and the ideal continuous passive-reflection environment.

For each BS, let
$\mathbf W_n\triangleq
[\mathbf w_{n,1},\ldots,\mathbf w_{n,N_T^{BS}}]$.
Using the Cauchy--Schwarz inequality and the per-BS power
constraint gives $
p_n(\mathcal L)
\triangleq
\left\|
\mathbf W_n^H\mathbf h_n(\mathcal L)
\right\|_2^2
\leq
\left\|\mathbf W_n\right\|_F^2
\left\|\mathbf h_n(\mathcal L)\right\|_2^2
\leq
P_{\max}\Gamma_R(N^R)$. Using the concavity of the perspective function
$B\log_2(1+x/B)$ under
$\sum_{n=1}^{N^{BS}}B_n\leq B_{\max}$ then yields
\begin{eqnarray}
C_T(\mathcal L)
\leq
\frac{B_{\max}}{|\mathcal A|}
\log_2
\left(
1+
\frac{
N^{BS}P_{\max}\Gamma_R(N^R)
}{
\sigma^2B_{\max}
}
\right).
\end{eqnarray}
Integration over $\mathcal A$ completes the proof.

\section{Proof of Mismatch Scaling}
\label{appendix:mismatch_scaling}

This appendix proves Theorem~\ref{thm:mismatch_scaling_new}. We first bound the JS divergence by the branch-wise $L^2$ approximation errors and establish the BS/RIS resolution hierarchy. We then derive the RIS-side rates for spatially diffuse and hotspot-dominated residuals. For fixed $N^{BS}$ and $N^R$, the mismatch scaling below refers to the best achievable value over the admissible ADD-RBF capacity-density family. Here, $\sigma_{\min}(\mathbf M)$ denotes the smallest singular value of $\mathbf M$, and $a_N\asymp b_N$ means that their ratio is bounded above and below by positive constants independent of $N^{BS}$ and $N^R$.

\subsection{Common Error Reduction and BS/RIS Spatial Resolution}

For non-negative normalized densities $f_c$ and $f_d$, the pointwise JS integrand with base-two logarithms is upper bounded by $\frac{1}{2}|f_c(\mathcal L)-f_d(\mathcal L)|$. Hence, integration over $\mathcal A$ and the Cauchy--Schwarz inequality give
\begin{small}
\begin{eqnarray}
\xi(C_S,D_T)
\leq
\frac{1}{2}\Vert f_c-f_d\Vert_{L^1(\mathcal A)}
\leq
\frac{|\mathcal A|^{1/2}}{2}
\Vert f_c-f_d\Vert_{L^2(\mathcal A)}.
\label{eqn:JS_L2_bound_balanced_appendix}
\end{eqnarray}
\end{small}

Decompose the normalized traffic density into non-negative components as $f_d=f_g+f_\ell$, where $f_g$ is the large-scale component assigned to the BS branch and $f_\ell$ is the remaining local component. Choose non-negative BS- and RIS-branch approximants $\widehat f_g$ and $\widehat f_\ell^R$ whose component masses are preserved by the corresponding amplitude coefficients, so that $\widehat f_c=\widehat f_g+\widehat f_\ell^R$ is normalized. Here, the admissible component families contain only approximants whose non-negative amplitudes are realizable through the bandwidth, beamforming, illumination, and normalization variables of the ADD-RBF model. Let $\mathcal V_B^g(N^{BS})$ and $\mathcal V_R^\ell(N^R)$ denote these admissible component families and define
\begin{eqnarray}
\mathcal E_g^B(N^{BS})
&\triangleq&
\inf_{\widehat f_g\in\mathcal V_B^g(N^{BS})}
\Vert f_g-\widehat f_g\Vert_{L^2(\mathcal A)},
\nonumber\\
\mathcal E_\ell^R(N^R)
&\triangleq&
\inf_{\widehat f_\ell^R\in\mathcal V_R^\ell(N^R)}
\Vert f_\ell-\widehat f_\ell^R\Vert_{L^2(\mathcal A)}.
\nonumber
\end{eqnarray}
For any such pair, the triangle inequality and \eqref{eqn:JS_L2_bound_balanced_appendix} give
\begin{eqnarray}
\xi(\widehat f_c,f_d)
\leq
\frac{|\mathcal A|^{1/2}}{2}
\left(
\Vert f_g-\widehat f_g\Vert_{L^2(\mathcal A)}
+
\Vert f_\ell-\widehat f_\ell^R\Vert_{L^2(\mathcal A)}
\right).
\nonumber
\end{eqnarray}
Let $\mathcal F_{\rm ADD}$ denote the admissible ADD-RBF capacity-density family for the specified values of $N^{BS}$ and $N^R$. Since the above branch-separable construction is a subfamily of $\mathcal F_{\rm ADD}$, taking the infimum yields the following upper bound on the best-achievable mismatch:
\begin{eqnarray}
\inf_{C_S\in\mathcal F_{\rm ADD}}
\xi(C_S,D_T)
=
\mathcal O \left(
\mathcal E_g^B(N^{BS})+
\mathcal E_\ell^R(N^R)
\right).
\label{eqn:JS_branch_error_balanced_appendix}
\end{eqnarray}

For the finite-smoothness results, assume that the admissible optimized atom centers contain quasi-uniform sequences over their effective supports, with mesh ratios bounded independently of $N^{BS}$ and $N^R$. The BS fill distance over the effective support of $f_g$ satisfies $h_B^g=\mathcal O((N^{BS})^{-1/2})$. If $f_g$ belongs to the finite-smoothness approximation class of the BS direct-path dictionary with order $t_g$, the standard scattered-data estimate gives \cite{wendland2004scattered,narcowich2006sobolev}
\begin{eqnarray}
\mathcal E_g^B(N^{BS})
=
\mathcal O \left((h_B^g)^{t_g}\right)
=
\mathcal O \left((N^{BS})^{-t_g/2}\right).
\label{eqn:BS_error_balanced_appendix}
\end{eqnarray}

The above branch-wise approximation is also consistent with the different spatial scales induced by the BS direct-path and RIS-assisted atoms.
For the BS branch, the direct-path power atom is proportional to $K_B(d)
= \left(\gamma d^\alpha+\beta\right)^{-2}$, where $d$ denotes the BS--user distance. The derivative of $K_B(d)$ with respect to $d$ is
$K_B'(d)
\triangleq
\frac{\mathrm d K_B(d)}{\mathrm d d}
=
-2\alpha\gamma d^{\alpha-1}
\left(\gamma d^\alpha+\beta\right)^{-3}$.
Accordingly, its local variation length is defined as
\begin{eqnarray}
\ell_B(d)
\triangleq
\left|
\frac{K_B(d)}{K_B'(d)}
\right|=
\frac{\gamma d^\alpha+\beta}
{2\alpha\gamma d^{\alpha-1}},
\nonumber
\end{eqnarray}
which characterizes the spatial distance over which the direct-path atom undergoes an appreciable relative variation. We summarize the effective BS-side spatial width over the service region by
\begin{eqnarray}
\bar\ell_B
\triangleq
\iint_{\mathcal A}f_d(\mathcal L)
\min_{1\leq n\leq N^{BS}}
\ell_B\left(
\left\|
\mathcal L-\mathcal L_n^B
\right\|
\right)
\mathrm d\mathcal L.
\label{eqn:effective_BS_width_balanced_appendix}
\end{eqnarray}

For an effective BS--RIS pair, the reflected-power atom further contains a phase-controlled array factor. For a target location $\mathcal L_0$, define the residual outgoing phase mismatch as
$\boldsymbol\Delta_{n,m}(\mathcal L;\mathcal L_0)
=\frac{2\pi d_R}{\lambda_c}
[\mathbf u_m^u(\mathcal L)-\mathbf u_m^u(\mathcal L_0)]$,
and let $\mathbf J_{\Delta,n,m}$ denote its Jacobian with respect to $\mathcal L$. For an RIS employing a uniform rectangular array (URA), the normalized array factor is a product of two Dirichlet kernels, whose main-lobe widths are of orders $(N_x^R)^{-1}$ and $(N_y^R)^{-1}$ in the two phase-mismatch coordinates. After linearization around $\mathcal L_0$, the corresponding RIS-assisted spatial width is therefore
\begin{eqnarray}
\ell_{R,n,m}^{\Theta}(\mathcal L_0)
\asymp
\frac{1}{\sigma_{\min}\left(
\mathbf D_R\mathbf J_{\Delta,n,m}(\mathcal L_0)
\right)},
\,
\mathbf D_R=\operatorname{diag}(N_x^R,N_y^R).
\nonumber
\end{eqnarray}
For a nearly square URA, $N_x^R\asymp N_y^R\asymp\sqrt{N_T^R}$, and the traffic-weighted effective RIS width satisfies
\begin{eqnarray}
\ell_R^\Theta
\asymp
\frac{1}{
\sqrt{N_T^R} \overline\sigma_\Delta},
\label{eqn:RIS_width_balanced_appendix}
\end{eqnarray}
where $\overline\sigma_\Delta$ summarizes the angular sensitivity over the effective illuminated BS--RIS pairs and service region.

These two spatial scales explain the branch-wise construction used in Theorem~\ref{thm:mismatch_scaling_new}. Since an RIS is passive and relies on the incident BS field, the large-scale active-capacity footprint is primarily established by the BS direct-path branch. Once this footprint has been formed, with sufficient $N_T^R$, the hierarchy
$\ell_R^\Theta\ll\bar\ell_B$ allows the finer RIS-assisted atoms to resolve the remaining local variations within the illuminated region. Therefore, assigning $f_g$ to the BS branch and $f_\ell$ to the RIS branch gives a resolution-consistent achievable decomposition for the mismatch upper bound.

\subsection{Spatially Diffuse Residual: Coverage-Limited Approximation}

For a spatially diffuse residual, the unmatched local variation remains distributed over an extended two-dimensional support $\Omega_{\rm dif}$. Under quasi-uniform deployment, the RIS fill distance satisfies $h_{R,{\rm dif}}=\mathcal O((N^R)^{-1/2})$. If the diffuse component belongs to the finite-smoothness approximation class of the local RIS dictionary with order $t_\ell$, the scattered-data approximation estimate gives
\begin{eqnarray}
\mathcal E_\ell^R(N^R)
=
\mathcal O \left(h_{R,{\rm dif}}^{t_\ell}\right)
=
\mathcal O \left((N^R)^{-t_\ell/2}\right).
\label{eqn:diffuse_error_balanced_appendix}
\end{eqnarray}

\subsection{Hotspot-Dominated Residual: Fixed-Support Structured Refinement}

For a hotspot-dominated residual, let $f_\ell=\sum_{k=1}^{K}f_{\ell,k}$, where $K$ and the compact effective support of each hotspot component are independent of $N^R$. Additional RISs then refine the capacity profiles over the same fixed hotspot regions rather than progressively covering an expanding support. For a candidate RIS location $\boldsymbol z$ serving hotspot $k$, denote the corresponding normalized physical RIS-assisted atom by
\begin{eqnarray}
\varphi_k(\mathcal L,\boldsymbol z)
\propto
\frac{|A_k(\mathcal L,\boldsymbol z)|^2}
{L(\boldsymbol z,\mathcal L_{n_k}^B)^2
 L(\mathcal L,\boldsymbol z)^2},
\nonumber
\end{eqnarray}
where $n_k$ is an effective illuminating BS. Continuously varying the RIS location and phase configuration generates a spatial family of such physical atoms over each hotspot region, represented by the following synthesis.

\begin{Assump}[Structured Hotspot RIS Synthesis]
\label{assump:continuous_RIS_synthesis}
For each hotspot component $f_{\ell,k}$, there exist a fixed reference square $\mathcal U=[-1,1]^2$, an admissible RIS-location map $\boldsymbol\chi_k:\mathcal U\rightarrow\mathbb R^2$, and a non-negative bounded synthesis density $a_k$ such that
\begin{eqnarray}
f_{\ell,k}(\mathcal L)
&=&
\iint_{\mathcal U}
G_k(\mathcal L,\boldsymbol u)
\,\mathrm d\boldsymbol u,
\nonumber\\
G_k(\mathcal L,\boldsymbol u)
&\triangleq&
a_k(\boldsymbol u)
\varphi_k \left(
\mathcal L,\boldsymbol\chi_k(\boldsymbol u)
\right).
\nonumber
\end{eqnarray}
Any Jacobian determinant associated with $\boldsymbol\chi_k$ is absorbed into $a_k$. Moreover, $\boldsymbol u\mapsto G_k(\cdot,\boldsymbol u)$, viewed as an $L^2(\mathcal A)$-valued function, admits an analytic extension to a fixed product of Bernstein ellipses and is uniformly bounded there by a constant independent of $N^R$.
\end{Assump}

This sufficient regularity condition is motivated by the fixed hotspot supports and admissible RIS regions considered here: away from path-loss singularities, the far-field path-loss and steering factors vary regularly with the RIS location.

\begin{Lem}[Two-Dimensional Analytic RIS-Synthesis Discretization]
\label{lem:analytic_RIS_quadrature_balanced}
Under Assumption~\ref{assump:continuous_RIS_synthesis}, let $\mathsf{GL}_p$ be the $p$-point Gauss--Legendre rule on $[-1,1]$, with nodes $\{u_{p,i}\}_{i=1}^{p}$ and positive weights $\{\varpi_{p,i}\}_{i=1}^{p}$, and define
\begin{eqnarray}
\mathsf{GL}_p^{\otimes2}G_k
\triangleq
\sum_{i=1}^{p}\sum_{j=1}^{p}
\varpi_{p,i}\varpi_{p,j}
G_k \left(\cdot,(u_{p,i},u_{p,j})\right).
\nonumber
\end{eqnarray}
Then, there exist constants $C_k,b_k>0$, independent of $p$, such that
\begin{eqnarray}
\left\|
f_{\ell,k}-\mathsf{GL}_p^{\otimes2}G_k
\right\|_{L^2(\mathcal A)}
\leq
C_k\exp(-b_kp).
\label{eqn:quadrature_rate_balanced_appendix}
\end{eqnarray}
\end{Lem}
\IEEEproof
The analytic extension in Assumption~\ref{assump:continuous_RIS_synthesis} gives an $L^2(\mathcal A)$-valued tensor-product polynomial approximation of degree at most $2p-1$ in each coordinate with error $\mathcal O(\exp(-b_kp))$ \cite{zhao2013sharp}. The tensor-product Gauss--Legendre rule is exact for this polynomial, and the positivity and fixed sum of its weights make both the integral and quadrature operators uniformly bounded. Applying both operators to the approximation residual proves \eqref{eqn:quadrature_rate_balanced_appendix}.
\endIEEEproof

Assign $N_k^R$ RIS-assisted atoms to hotspot $k$ and set $p_k=\lfloor\sqrt{N_k^R}\rfloor$ \cite{bungartz2004sparse}. The tensor-product rule $\mathsf{GL}_{p_k}^{\otimes2}G_k$ uses $p_k^2\leq N_k^R$ admissible physical atoms: each quadrature node determines an RIS location, while its positive weight and $a_k$ determine the non-negative amplitude. Lemma~\ref{lem:analytic_RIS_quadrature_balanced} and $p_k\geq\sqrt{N_k^R}/2$ for $N_k^R\geq4$ yield
\begin{eqnarray}
\inf_{\widehat f_{\ell,k}^R}
\left\|
f_{\ell,k}-\widehat f_{\ell,k}^R
\right\|_{L^2(\mathcal A)}
\leq
C_k\exp \left(-\widetilde b_k\sqrt{N_k^R}\right),
\label{eqn:hotspot_component_rate_balanced_appendix}
\end{eqnarray}
where $\widetilde b_k>0$ is independent of $N_k^R$.

Because the number of hotspots is fixed, suppose that the optimized allocation satisfies $\sum_{k=1}^{K}N_k^R=N^R$ and $N_k^R\geq\omega_kN^R-1$, where $\omega_k>0$, $\sum_k\omega_k=1$, and $\{\omega_k\}$ are independent of $N^R$. Summing \eqref{eqn:hotspot_component_rate_balanced_appendix} over the fixed hotspot set gives
\begin{eqnarray}
\mathcal E_\ell^R(N^R)
=
\mathcal O \left(
\exp[-c_R\sqrt{N^R}]
\right),
\label{eqn:hotspot_error_balanced_appendix}
\end{eqnarray}
where $c_R>0$ is independent of $N^R$. The square root in the exponent follows from the exponential convergence in the per-coordinate quadrature order and the two-dimensional relation $N_k^R\asymp p_k^2$.

Finally, substituting \eqref{eqn:BS_error_balanced_appendix} and \eqref{eqn:diffuse_error_balanced_appendix} into \eqref{eqn:JS_branch_error_balanced_appendix} gives the best-achievable diffuse-residual mismatch
$\mathcal O((N^{BS})^{-t_g/2}+(N^R)^{-t_\ell/2})$.
Under Assumption~\ref{assump:continuous_RIS_synthesis}, substituting \eqref{eqn:hotspot_error_balanced_appendix} instead gives the best-achievable hotspot-dominated mismatch
$\mathcal O((N^{BS})^{-t_g/2}+\exp[-c_R\sqrt{N^R}])$, completing the proof.

\end{appendices}

\bibliographystyle{IEEEtran}
\bibliography{IEEEfull}

\end{document}